\documentclass[journal,twocolumn,10pt]{IEEEtran}

\normalsize

\usepackage{cite, amsmath,amsfonts,amsthm,amssymb}
\usepackage{algorithmic,algorithm,float,bbm}
\usepackage{multirow,balance,hyperref}
\usepackage{enumerate,bbding}
\usepackage{epsfig,graphicx,subfigure}

\theoremstyle{plain}

\theoremstyle{definition}
\newtheorem{defn}{Definition}

\theoremstyle{remark}

\usepackage{booktabs}
\usepackage{threeparttable}

\graphicspath{{images/}}

\begin{document}

\title{Virtual Reality: A Survey of Enabling Technologies and its Applications in IoT}
\author{Miao Hu, Xianzhuo Luo, Jiawen Chen, Young Choon Lee, Yipeng Zhou, Di Wu
\thanks{M. Hu, X. Luo, J. Chen, and D. Wu are with School of Computer Science and Engineering, Sun Yat-sen University, Guangzhou 510275, China, and also with Guangdong Key Laboratory of Big Data Analysis and Processing, Guangzhou 510006, China.}
\thanks{Y.C. Lee and Y. Zhou are with Department of Computing, Faculty of Science and Engineering, Macquarie University, Sydney, NSW 2109, Australia.}
}

\maketitle

\begin{abstract}
Virtual Reality (VR) has shown great potential to revolutionize the market by providing users immersive experiences with freedom of movement. Compared to traditional video streaming, VR is with ultra high-definition and dynamically changes with users' head and eye movements, which poses significant challenges for the realization of such potential. In this paper, we provide a detailed and systematic survey of enabling technologies of virtual reality and its applications in
Internet of Things (IoT). We identify major challenges of virtual reality on system design, view prediction, computation, streaming, and quality of experience evaluation. We discuss each of them by extensively surveying and reviewing related papers in the recent years. We also introduce several use cases of VR for IoT. Last, issues and future research directions are also identified and discussed.
\end{abstract}


\maketitle

\section{Introduction}
The Internet of Things (IoT) enables devices to share data and information so as to provide added convenience and control for users and even allow users to automate simple processes. Tens of billions of connected IoT devices already exist in the world and this number is anticipated to keep increasing as the Internet connectivity has become a standard feature for a great number of electronics devices. The size of the global IoT market has grown to 212 billion US dollars by 2019. The IoT market revenue is forecasted to reach 1.6 trillion by 2025 \cite{Statista2019IoT}.

Virtual reality (VR) enables IoT users to obtain extraordinary immersive experience through visible display with multiple dimensions, where users can freely change view angles to see what they want. Many companies (e.g., {Facebook}, {YouTube} and {Periscope}) have built platforms to provide VR services, and have gained wide attention around the world. Today, the world's largest corporations are all investing heavily in the VR development, including {Adobe}, {Apple}, {Amazon}, {Google}, {Microsoft} and {Samsung}. Even the US army is beginning to use VR technology to train soldiers. This survey aims to answer how new trends and technologies in VR are being applied in real-world IoT applications.

With full freedom of movement, VR provides immersive user experiences for IoT applications. To support the dynamic viewpoint-changing characteristics, a VR system has to deliver multiple screens, which results in tremendous consumption of storage space and network bandwidth \cite{Afzal2017Characterization, Xiao2017MM_OpTile, Liu2019TMM}. VR for IoT also poses new challenges on system design, dynamic viewpoint prediction, complex encoding/decoding with high rendering cost, adaptive streaming, and effective quality of experience (QoE) evaluation, as follows.
\begin{itemize}
    \item \textbf{Efficient VR system design}. Compared to the standard streaming system, the VR system design faces new challenges, especially in the scenarios with limited communication and computation resources. When the cellular or WiFi is adopted for VR system design, it can become a bottleneck for content transmission, because the available wireless bandwidth is generally not enough for delivering the ultra high-definition (UHD) VR content, which might cause playback delay. Moreover, in VR video processing, the device needs extra time to render graphics and generate a full-motion-video playback version that works in real time. Compared to traditional video streaming, rendering on VR videos costs much more computation resources, especially for rendering on the object boundary. How to establish an efficient VR system for real-time applications is of the first importance.
    \item \textbf{Dynamic user view prediction}. The user viewpoint is dynamic, which will be affected by both frame property and user behavior. It is the content on the frame that attracts users to swing heads. Thus, it is essential to extract key features from each video frame. \emph{Second}, different users show non-identical behaviors on watching the same frame. User QoE will significantly deteriorate if the screen in the field of view (FoV) is delivered with a low bitrate due to the inaccurate prediction of FoV.
    \item \textbf{High computation requirement}. In real-time IoT applications, the VR system design faces many challenges on complex computation operations, including encoding/decoding and rendering. Compared to the standard video, the higher resolution of the VR video makes the projection and coding/decoding process more complicated, thus introducing a longer processing time. In general, a practical VR system has to leverage compression, which can reduce the size by 90\% on average at the cost of millisecond-level decoding latency per frame.
    \item \textbf{Adaptive VR streaming}. The large size of uncompressed VR scenes makes directly storing or streaming VR content over bandwidth-limited wireless radio infeasible. This motivates the design of a relatively small buffer for caching, leaving a great challenge to the adaptive VR streaming design.
    \item \textbf{Reasonable QoE evaluation}. The resolution of VR videos is commonly UHD, such as 4K, 8K~\footnote{4K and 8K are four and eight times the pixel resolution, respectively.}. If all frames are delivered with UHD, the bandwidth consumption will be unaffordable. To mitigate the surge of bandwidth consumption, it is crucial to predict users' FoV, based on which we only need to deliver screens falling in the FoV. However, perfectly evaluating users' quality of experience is quite challenging in practical IoT applications with diverse user view behaviors.
\end{itemize}

In this survey, these challenges will be further discussed from Section \ref{sec_archi} to Section \ref{sec_stream} one by one.
A typical VR appliacation process mainly includes: user head movement, content generation (including projection, coding, rendering), delivery, display, and user experience. The above-mentioned issues put more urgent challenges on each processing component of the VR systems. We hereby summarize enabling technologies for constructing VR systems.

The reminder of this survey is organized as follows.
Section~\ref{sec_archi} lists popular architectures of VR systems.
Section~\ref{sec_HMD} describes characteristics of commercial VR devices. Moreover, the FoV prediction methods are also illustrated.
Section~\ref{sec_project} summarizes the projection and coding methods for VR streaming.
Section~\ref{sec_stream} surveys the adaptive streaming strategies and enumerates the QoE metrics for evaluating the VR streaming performance.
Section~\ref{sec_app} presents IoT use cases of utilizing VR technologies.
Section~\ref{sec_issue} discusses challenging issues and outlooks future research directions.
Finally, Section~\ref{sec_conclusion} concludes the work of this survey.


\begin{figure*}
\centering
  \includegraphics[width=0.85\textwidth]{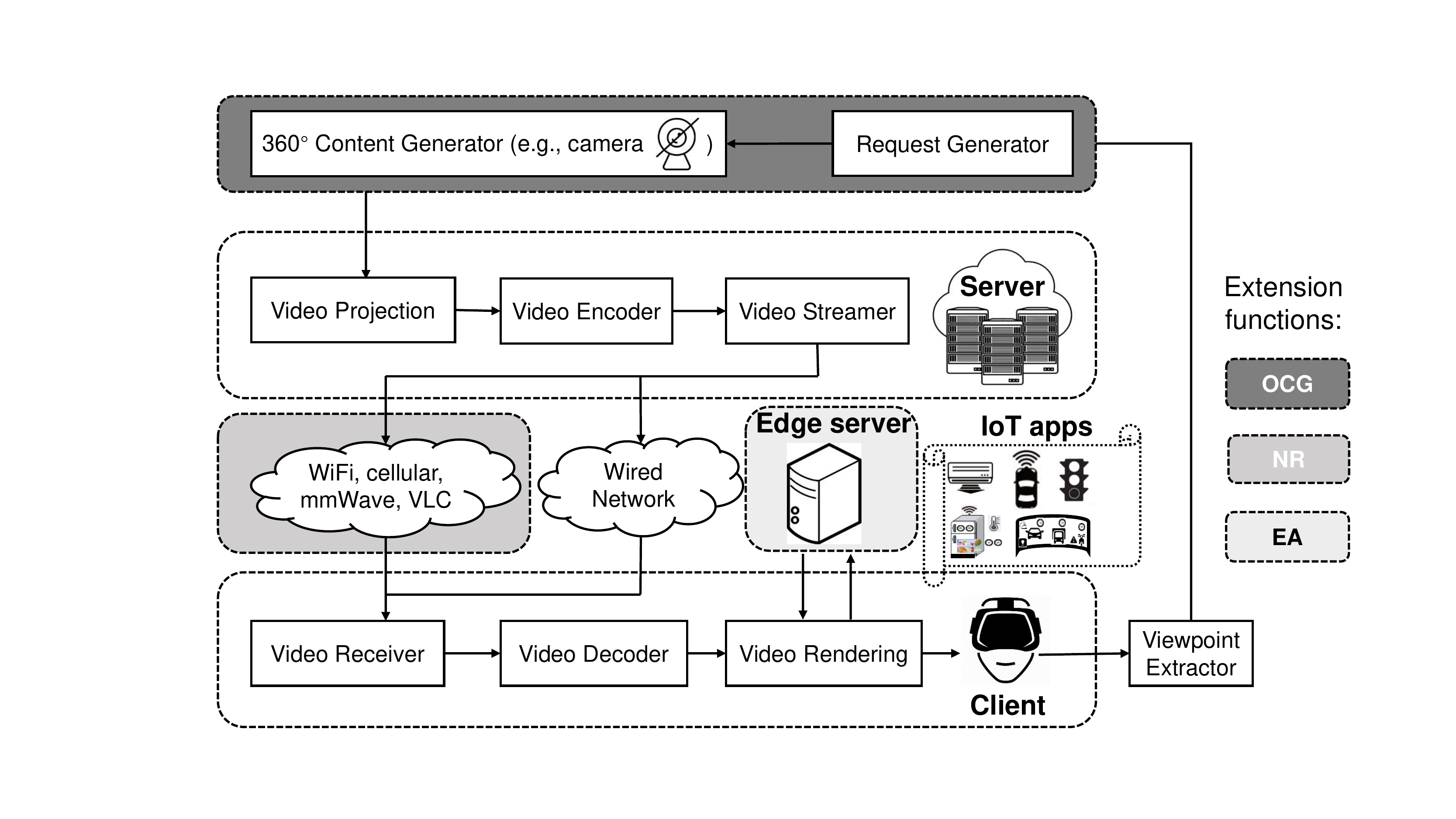}\\
  \caption{A general architecture and extensions of the systems for virtual reality in IoT applications.}\label{arch}
\end{figure*}

\section{Architecture of VR Systems}\label{sec_archi}
This section discusses several types of existing VR systems in the context of their design. Different solutions emphasize on various aspects and have distinct optimization objectives for different applications. In particular, we overview various VR systems, describe their components and features, and list some typical examples in each type.

\subsection{Standard Client-Server (CS) Architecture}
In the client-server architecture (Figure \ref{arch}), the server component provides a function or service to one or more clients, which initiate requests for such services. Such architecture is relatively common, which covers a wide range of VR systems in the literature.
For example, \cite{Xie2017MM_360ProbDASH} proposed 360ProbDASH, which leverages a probabilistic model of viewport prediction to predict a viewer's orientation and uses a target-buffer-based rate control algorithm to perform rate adaptation in the client. The server crops video frames into spatial tiles and encodes the tiles into multi-bitrate segments.

However, the adoption of the standard CS architecture for VR systems is hindered due to the following reasons:
\begin{itemize}
    \item \textbf{Live streaming adaption}. Since the standard CS architecture is generally applied to on-demand IoT services, it needs additional efforts like video generation and projection modules, to process 360-degree vision generated in real time when supporting VR streaming.
    \item \textbf{Transmission bandwidth limitation}. In the standard CS architecture, when the cellular or WiFi is adopted, it can become a bottleneck for content transmission, because the available wireless bandwidth is generally not enough for delivering the UHD VR content, which might cause playback delay. In most IoT applications, e.g., IoV, this limitation might cause serious safety consequences.
    \item \textbf{Urgent real-time requirement}. In the standard CS architecture, videos are streamed directly over the network to the client, during which the video is rendered on the client. This might cause a long processing latency and serious heat radiation. By leveraging edge servers to render or cache desired content close to users, we can further reduce network latency and enhance real-time experience.
\end{itemize}

In the following, we list some representative solutions to the above challenges.

\subsection{Live VR System}
With the emerging mobile omni-directional cameras and untethered head-mounted displays, it is likely for live VR video to become reality, which needs streaming VR videos in real time.
In comparison to the standard CS architecture, the live VR system adds one new function, live streaming generator as shown in Figure \ref{arch}.
For example, in VR system design, the three-dimensional visible navigation service can be implemented in two ways, i.e., \emph{on-demand stored content} and \emph{real-time content} generation.
For on-demand content provision, the visible content was generated for all possible cases in advance. This introduces high cost on content generation and storage for the application provider.
While for the live VR system, the implementation might be with less burden, but will introduce the delay on data preprocessing and video rendering.
\cite{Shi2019MMSys} implemented a mobile VR system that uses an edge cloud to stream live VR content to mobile devices using LTE networks.

\subsection{New Radio (NR)-driven VR System}
Due to the high data rate requirement of VR systems, wireless communication links (e.g., cellular and WiFi) have become a bottleneck \cite{Qian2016AllThingsCellular, Mangiante2017VRARNetwork, Sun2018MMSys}. Therefore, it is essential to optimize VR streaming by introducing new radio (Figure \ref{arch}). In comparison to the standard CS architecture, the NR-driven system adds new radio that serves as communication links, to achieve breakthrough especially in the indoor transmission.

Several studies have used mmWave links for VR \cite{Abari2017NSDI, Zhong2017APSys}. Typically, mmWave can support the transmission of uncompressed VR frames for a single user, which eliminates decompression latency at the client. However, a key challenge in using mmWave links for VR applications is that they may be easily blocked by a small obstacle, such as the player's hand.
Compared to mmWave, visible light communication can supplement a higher downlink data rate, which matches well with the high data volume requirement for VR streaming \cite{Khan2019arXiv}. However, the practical downlink rate will be seriously affected by the accuracy of indoor positioning, without which the download performance cannot be guaranteed and thus the user view QoE will be deteriorated.

\subsection{Edge-accelerated VR System}
Because offloading the application to the cloud is not always an option due to the high and often unpredictable network latencies, the concept of edge has been introduced: nearby infrastructure offering virtual machines for remote execution \cite{HU2017JNCA, HSIEH2018JNCA, RAY2019JNCA, ELAZHARY2019JNCA, LIU2020JNCA}.

As illustrated in Figure \ref{arch}, utilizing the emerging edge computing paradigm in VR systems can reduce power consumption and computation load for user end devices and improve responsiveness of the IoT service.
\cite{Zhang2019NOSSDAV} optimized the video generation stage of the edge-based VR services by using the available edge computing resources for renderring and encoding. Towards optimally placing video rendering and encoding at the edge, the proposed task assignment scheme is proved to be superior.
Some work proposed to reduce network latency and bandwidth demands by caching popular content close to the users. \cite{Mahzari2018MM} presented an FoV-aware caching policy to improve caching performance at the edge server.
\cite{Hou2018VRARNetwork} also utilized the user’s mobile device as an edge computing node, in which case the additional challenge of having to do predictive view generation in the mobile device will need to be addressed.

\subsection{Parallel VR System}
To achieve a lower latency, \cite{Liu2018MobiSys} employed a parallel rendering and streaming system to reduce the add-on streaming latency, by pipelining the rendering, encoding, transmission and decoding procedures. The proposed system consists of two parts: simultaneous rendering and encoding, and parallel streaming. Moreover, they built a parallel frame rendering and streaming pipeline to reduce the streaming latency.
Similarly, \cite{Duanmu2017JSTSP} proposed a framework that consists of two tiers of video representations. In the two-tier VR system, each segment is coded as a base-tier chunk, and multiple enhancement-tier chunks.
The base-tier chunks represent the entire 360 view at a low bit rate and the enhancement-tier chunks represent the partial 360 view at a high bit rate.
Unlike most of the previous systems, the two-tier scheme simultaneously addresses the accommodation of unexpected head motions and network variations, to provide users with a stable immersive experience.
\cite{DIASDEASSUNCAO2018JNCA} surveyed stream processing engines and mechanisms for exploiting resource elasticity features in distributed stream processing.

\subsection{Collaborative VR System}
Collaborative VR systems allow multiple users to share a real world environment, including computer-generated images in real time. Collaborative VR systems require a system design that takes into account scalability issues. \cite{Fernandez2014JNCA} presented the experimental characterization of collaborative VR systems based on mobile phones, providing quantitative results about well-known performance metrics in distributed VR systems.

Network softwarization paradigms promise numerous improvements regarding the scalability, flexibility, as well as resource and cost efficiency of the systems. In software defined network (SDN)-enabled networks, the logically centralized control and the separation of control and data planes pave the way for programmable networks \cite{Jarschel2014CM}. With virtualized network functions (NFV), dedicated hardware middleboxes are replaced with software instances. Network softwarization enables adaptability to fluctuations in terms of service-level requirements in the context of service function chaining (SFC) \cite{Lange2019CNSM}.

\section{HMD and FoV Prediction}\label{sec_HMD}
In this section, we first introduce the basic information of head-mounted displays (HMDs), and specify available HMD datasets. Furthermore, we summarize the state-of-the-art FoV prediction algorithms.

\subsection{Characteristics of HMDs}
\begin{table*}
\renewcommand\arraystretch{1.35}
\caption{Description of HMDs}
\label{HMD_table}
\centering
\linespread{1}\selectfont
\begin{tabular}{|c|c|c|c|c|c|}
\hline \textbf{Type}       &   \textbf{HMD}       & \textbf{DoF} & \textbf{Resolution}     & \textbf{Refresh Frequency} & \textbf{FoV}        \\ \hline\hline
\multirow{6}{*}{\textbf{Tethered VR}}   & \cite{PlayStationVR} & 6   & 960$\times$1080 pixels per eye   & 90/120Hz  &  $\approx 100^\circ$ \\ \cline{2-6}
    & HTC VIVE Cosmos & 6   & 1440$\times$1700   pixels per eye & 90Hz  & 110$^\circ$       \\ \cline{2-6}
    & HTC VIVE Pro & 6   & 1440$\times$1600   pixels per eye & 90Hz   & 110$^\circ$       \\ \cline{2-6}
    & HTC VIVE  & 6   & 1080$\times$1200  pixels per eye & 90Hz   & 110$^\circ$       \\ \cline{2-6}
    & Oculus Rift DK2 & 6   & 960$\times$1080 pixels per eye      & 75Hz  & 100$^\circ$       \\ \cline{2-6}
    & Oculus Rift S & 6   & 1440$\times$1280  pixels per eye & 80Hz    & $\approx 115^\circ$ \\ \hline
\multirow{3}{*}{\textbf{Untethered VR}} & HTC VIVE Focus & 6   & 2880$\times$1600  pixels   & 75Hz         & 110$^\circ$       \\ \cline{2-6}
    & Oculus Go      & 3   & 2560$\times$1440  pixels    & 60/72Hz  & $\approx 100^\circ$ \\ \cline{2-6}
    & Oculus Quest   & 6   & 1440$\times$1600  pixels per eye & 72Hz   & 100$^\circ$     \\ \hline
\multirow{2}{*}{\textbf{Untethered AR}}   & Microsoft HoloLens & 6   & 2048$\times$1080 pixels  &  240Hz &  $120^\circ$ \\ \cline{2-6}
    & Google Glass & 6   & 640$\times$360 pixels  &  N/A  & 80$^\circ$       \\ \hline
\end{tabular}
\vspace{-5pt}
\end{table*}

\begin{table*}
\renewcommand\arraystretch{1.35}
\caption{Characteristics of public datasets}
\label{character_dataset_table}
\centering
\begin{tabular}{|c|c|c|c|c|c|c|c|}
\hline
\textbf{Literature}                        & \textbf{HM} & \textbf{EM} & \textbf{Content} & \textbf{Analysis} & \textbf{Software} & \textbf{Sampling frequency} & \textbf{\# Users} \\ \hline\hline
\cite{Lo2017MMSys}        & \checkmark    &    & \checkmark     &      &       & 30Hz       & 50     \\ \hline
\cite{Corbillon2017MMSys} & \checkmark    &      &       & \checkmark     & \checkmark  & 30Hz   & 59 \\ \hline
\cite{Wu2017MMSys}        & \checkmark    &      &      & \checkmark    &        & N/A     & 48   \\ \hline
\cite{David2018MMSys}     & \checkmark    & \checkmark    &         & \checkmark &    & 250Hz   & 57 \\ \hline
\cite{Ozcinar2018QoMEX}  &    & \checkmark     &       & \checkmark    & \checkmark   & N/A  & 17  \\ \hline
\cite{Duanmu2018ICME}     &   & \checkmark     &      & \checkmark     & \checkmark & N/A     & 50  \\ \hline
\cite{Fremerey2018MMSys_AVTrack}  & \checkmark    &        &          & \checkmark  & \checkmark     & 200 Hz    & 48        \\ \hline
\cite{Nasrabadi2019MMSys} &        & \checkmark    &     & \checkmark   & \checkmark  & 60Hz   & 60  \\ \hline
\end{tabular}
\vspace{-5pt}
\end{table*}

A typical HMD is a display device that can playback VR content like 360-degree videos. The user can wear an HMD on the head, with sensors enabling head movement (HM) tracking, eye movement (EM) tracking, etc. While viewers' FoV reaches almost 180$^\circ$, usually HMDs can provide an FoV of around 100$^\circ$. Generally, an HMD can be categorized into a tethered or untethered device, depending on whether it needs a cable to connect to the computer. When wearing an HMD, viewers can move in six basic ways, including rotational movements around the $x$, $y$ and $z$ axes and translational movements along the axes, namely \emph{6 degrees of freedom} (\emph{DoF}). Some HMDs are able to track all the aforementioned movements, which mean they provide 6DoF, while others can only track the rotational movements and thus offer 3DoF.

Commercial HMDs include Oculus Rift DK2, Oculus Rift CV1, HTC Vive, Microsoft HoloLens, Google Glass and so on, most of which are connected to a PC workstation. We collect basic information of these commercial HMDs and make a summary in Table \ref{HMD_table}.

\subsection{Tracing Records on HMDs}
For research and system development, it is always required for recording the traces on HMDs. The measured traces can be categorized into: 1) head movement traces, 2) eye movement traces and 3) content traces, according to the specific IoT applications requirements.

\subsubsection*{Head Movement Tracing}
As a VR user can change viewport position by moving his or her head, head movement traces can be used to study a user's head motion pattern. The general tracing collection procedure is shown as follows.
\begin{itemize}
    \item Initially, when watching a VR video streamed to the HMD, viewers can look around to change their viewport position freely and get familar with the HMD.
    \item During the video playback, viewers' head movement data need to be recorded. \cite{Lo2017MMSys} used OpenTrack as the sensor logger to record the viewer orientation, while \cite{Corbillon2017MMSys} developed a software to log the viewer's head position at each frame.
    \item After recording, the head motion data is stored in .csv, .txt or some other format.
\end{itemize}

There have been many studies on the characteristics and statistics of the head movement traces. For example, \cite{Wu2017MMSys} visualized the paths of their viewers' gazing directions in 3D space, and analyzed the angular velocity of their head movement. \cite{David2018MMSys} analyzed fixation count per latitude and per longitude to provide some insights about 360-degree content exploration. On top of that, some researchers also considered measuring viewing experience. \cite{Fremerey2018MMSys_AVTrack} asked the viewers to fill in the simulator sickness questionnaire and analyzed the result.

\subsubsection*{Eye Movement Tracing}
Eye movement traces are essential to study users' gaze behaviors \cite{Rai2017MMSys}. However, compared to head movement traces, it is more difficult to collect eye movement traces, since some auxiliary equipment (e.g., eye-tracking camera) is usually needed. \cite{David2018MMSys} used an HTC VIVE headset equipped with an SMI (SensoMotoric Instrument) eye-tracker to display VR streaming and capture eye movement data. \cite{Xu2018CVPR} captured viewers' eye fixations with an HTC VIVE headset and a 7invensun a-Glass eye tracker (an in-helmet camera system). On top of that, some studies were conducted based on their eye tracking dataset, including consistency of gaze points, distribution of eye fixations in latitude and longitude, etc. Besides, some researchers use self-developed testbeds to replace the eye-tracking equipment. \cite{Ozcinar2018QoMEX} designed a testbed to gather viewer's viewport center trajectories when watching VR streaming without the need of specific eye tracking devices.

\begin{table*}
\renewcommand\arraystretch{1.35}
\caption{Specifications of public 360-degree videos}\label{specification_table}
\centering
\linespread{1}\selectfont
\begin{tabular}{|c|c|c|c|c|c|l|}
\hline
\textbf{Literature}   & \textbf{Video Length}  & \textbf{Frame Rate} & \textbf{Quantity} & \textbf{Resolution} & \textbf{Label Information}    \\ \hline\hline
\cite{Lo2017MMSys}        & 60s      & 30fps   & 10     & 4K    & Fast-paced/Slow-paced, CG/NI     \\ \hline
\cite{Corbillon2017MMSys} & 70s       & 25-60fps   & 5        & 3840$\times$2048                                                  & Tourism, Thrill,  Discovery                  \\ \hline
\cite{Wu2017MMSys}        & 2'44''-10'55''  & 25-30fps   & 18       & \begin{tabular}[c]{@{}c@{}}2160$\times$1080/2560$\times$1440/\\1920$\times$960/2880$\times$1440/\\3840$\times$1920/2400$\times$1200 \end{tabular} &         \begin{tabular}[c]{@{}c@{}}Performance, Sport, Film,\\ Documentary, Talkshow     \end{tabular}         \\ \hline
\cite{David2018MMSys}     & 20s       & 24-30fps   & 19       & 3840$\times$1920                                                  & \begin{tabular}[c]{@{}c@{}}Indoor/Outdoor, Rural/Natural,\\ Containing people faces, etc.\end{tabular}  \\ \hline
\cite{Ozcinar2018QoMEX}   & 10s       &  N/A      & 6        & 4K$\times$2K    &
\begin{tabular}[c]{@{}c@{}}Use spatial and temporal\\ indices to represent a  broad \\range of content complexities     \end{tabular}                    \\ \hline
\cite{Duanmu2018ICME}     & 60s-120s    &  N/A   & 12       & 2048$\times$1024/3840$\times$2048                                        &  \begin{tabular}[c]{@{}c@{}}Virtual tour, VR gaming, Stage,\\ Sports, Movies, etc. \end{tabular}  \\ \hline
\cite{Fremerey2018MMSys_AVTrack}  & 30s        & 30fps    & 20     & 4K   & N/A \\ \hline
\cite{Nasrabadi2019MMSys} & 60s    & 24-30fps   & 28       & 2560$\times$1440-3840$\times$2160                                        & Moving objects, Camera motion  \\ \hline
\end{tabular}
\vspace{5pt}
\end{table*}

\subsubsection*{Content Tracing}
As the content of VR videos and viewer's unique viewing habit affect a viewer's future viewport jointly, it is necessary to take content traces into consideration.
\cite{Lo2017MMSys} generated image saliency maps and motion maps for each 360-degree video they collected. With content traces, content-based algorithm can be applied to predict future viewport.
\cite{Fan2017NOSSDAV} proposed a fixation prediction network that concurrently leverages content and sensor traces.

In summary, Table II compares the characteristics of the three types of datasets. Table III specifies the main properties of the 360-degree videos used by the datasets.

\begin{table*}
\renewcommand\arraystretch{1.35}
	\caption{FoV Prediction}\label{FoV Prediction}
	\centering
    \linespread{1}\selectfont
		\begin{tabular}{|p{2cm}|p{1.5cm}|p{3cm}|p{2cm}|p{2cm}|p{2cm}|p{2.5cm}|}
			\hline \bf{Category} & \bf{Literature} & \bf{Solution} & \bf{Model} & \bf{Input} & \bf{Output} & \bf{Evaluation} \\ \hline
			\hline \multirow{5}{*}{\shortstack{User Driven\\FoV Prediction}} & Rubiks \cite{He2018MobiSys_Rubiks} & Used past head motion to predict future head motion based on head movement trace & LR & X and Y axis movement & X and Y axis movement & The average prediction errors along the X and Y axes are 25$^{\circ}$ and 8$^{\circ}$ \\
			\cline{2-7} & \cite{Hou2018VRARNetwork} & Used past head motion to predict future head motion based on head movement trace & LSTM & Viewpoint matrix in past 2s & The predicted probabilities for all possible tiles & LSTM achieves an FoV prediction accuracy of more than 95.0\% \\
			\cline{2-7} & \cite{Mahzari2018MM} & Used past FoV samples to predict future FoV & WLR & 10 most recent FoV samples & Predicted FoV of the next moment & The FoV prediction was regarded as a middle process without evaluation  \\
			\cline{2-7} & Flare \cite{Qian2018Flare} & Used past head position to predict the trajectory of future viewports and use different model in different prediction windows & LR, RR, SVR & Head positions in the most recent history window & 30 predicted viewport in the prediction windows & For the four prediction window sizes (0.2s, 0.5s, 1.0s, 3.0s), the median FoV accuracy are 90.5\%, 72.9\%, 58.2\%, and 35.2\%  \\
			\cline{2-7} & DRL360 \cite{Zhang2019INFOCOM_DRL360} & Used past viewport to predict future viewport & LSTM & The historical viewport matrix & Predicted viewport of the next chunk & The average FoV prediction precision is about 92\%  \\
			\hline \multirow{3}{*}{\shortstack{Content Driven\\FoV Prediction}} & \cite{Nguyen2018MM} & Integrated the panoramic saliency maps with user head orientation history & LSTM & Saliency map and head orientation map & Predicted head orientation of the next moment  & The FoV prediction accuracy is about 75\% \\
			\cline{2-7} & \cite{Fan2017NOSSDAV} & Leveraged sensor-related and content-related features to predict the viewer fixation in the future & LSTM & HMD orientations, saliency maps and motion maps & viewing probability of the next $n$ frames & The FoV prediction accuracy is 86.35\% \\
			\cline{2-7} & \cite{Lee2019NOSSDAV} & Integrated the information of human visual attention with the contents to deliver high-quality tiles to the ROI & CNN & Eye fxation information and video frames & The predicted saliency map & The saliency map prediction was regarded as a middle process without evaluation \\
			\hline \multirow{1}{*}{\shortstack{\\User-and-content\\Driven FoV\\Prediction}} & CLS \cite{Xie2018MM_CLS} & Integrated past fixations with cross-users' ROI found from their historical fixations to predict the viewport in the future & DBSCAN and SVM & Tile view probability and user's past fixations & Predicted viewport of the next moment & CLS can achieve 85\% and 90\% prediction precision in prediction time horizon of 3 seconds and 1 second respectively \\
			\hline
	\end{tabular}
\end{table*}

\subsection{FoV Prediction}
\begin{defn}
    \textbf{FoV} stands for \emph{field of view}, or \emph{field of vision}. The FoV in a VR system is how much of a given range the user can see on his/her HMD screen at once.
\end{defn}
Some of the best VR headsets, like the HTC Vive Pro and Samsung HMD Odyssey have an FoV of 110 degrees. The Oculus Rift and the lower-priced standalone HMD (head-mounted display, same thing as a headset) Oculus Go have an FoV of about 100 degrees.
The objective of predicting the FoV in the future image is to serve for building a more efficient VR streaming strategy.
It is essential but not easy to conduct seamless bitrate control, which has been affected by both the server-client bandwidth and the client-side buffer that stores tiles prior to being played out. In the following, we will introduce the state-of-the-art FoV prediction algorithms. The VR streaming strategies will be summarized in Section~\ref{sec_stream}.

\begin{figure}
\centering
  \includegraphics[width=0.45\textwidth]{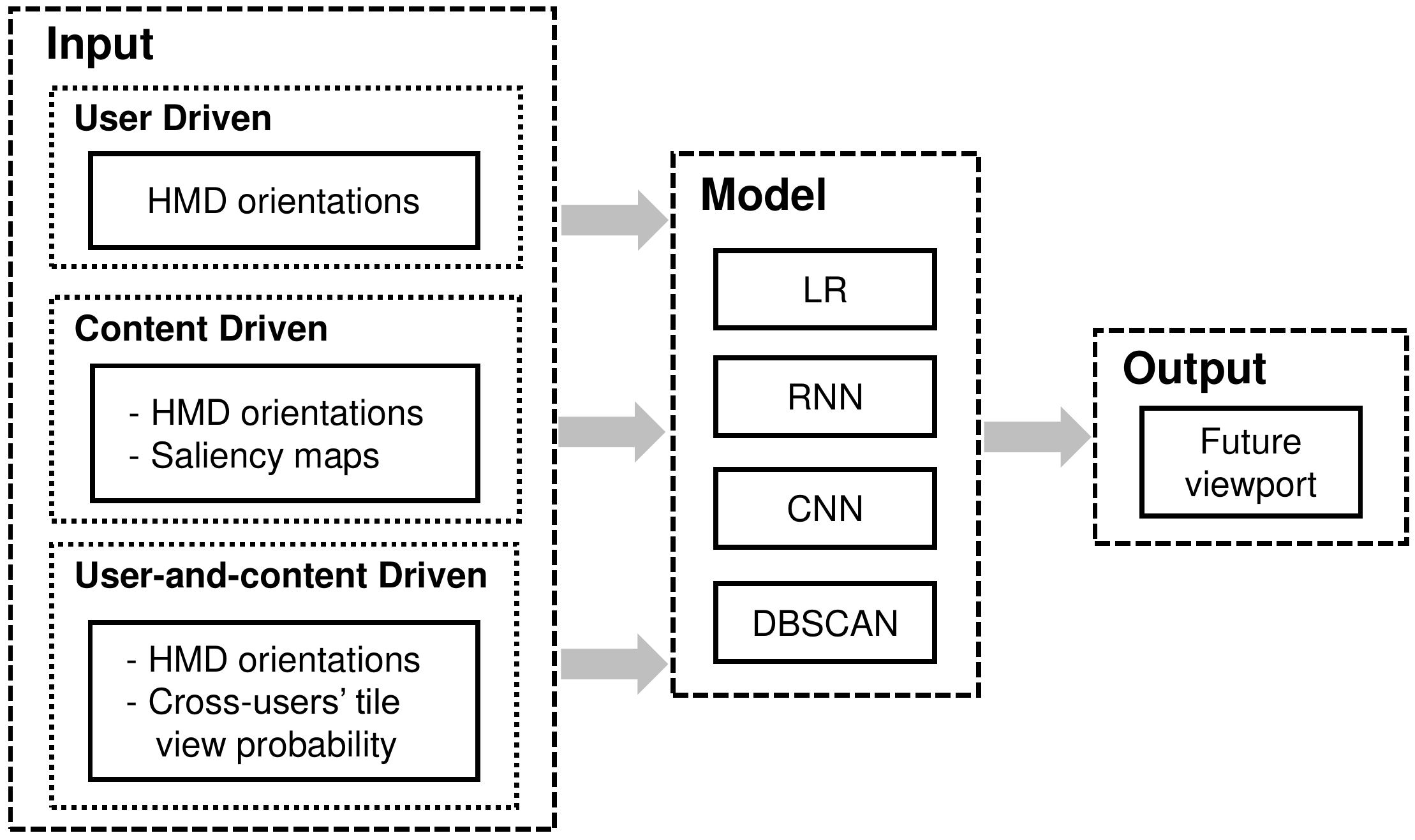}\\
  \caption{A general classification on FoV prediction in VR applications.}
  \label{FoVpredict}
\end{figure}

As shown in Figure \ref{FoVpredict}, based on the input data, the FoV prediction algorithm for VR applications can be generally classified into three classes: 1) user-driven algorithms, 2) content-driven algorithms, and 3) a hybrid of user-driven and content-driven algorithms. The three kinds of FoV prediction solutions are presented as follows.

\subsubsection*{User-driven FoV Prediction}
The user driven solution indicates that the past user head movement traces are used in predicting the future FoV.
Based on the utilized prediction methods, it can be further split into two kinds of models.
\begin{itemize}
    \item \emph{Linear regression (LR)}. \cite{He2018MobiSys_Rubiks} used past head motion to predict future head motion based on linear regression \cite{LR_DeepLearning}. Except LR, \cite{Qian2018Flare} also used ridge regression (RR) and support vector regression (SVR) with different prediction windows to predict future FoV.
    \item \emph{Long short-term memory (LSTM)}. \cite{Hou2018VRARNetwork} took the viewpoint matrix in past two seconds as input of the LSTM model and predicted the probabilities over FoV.
    \cite{Zhang2019INFOCOM_DRL360} compared the LSTM model for FoV prediction with LR algorithm, image-based CNN model \cite{Fan2017NOSSDAV} and KNN algorithm \cite{Xie2018MM_CLS}. The comparison result showed that the LSTM model performs the best.
\end{itemize}

\subsubsection*{Content-driven FoV Prediction}
The characteristics of the content can also help predict FoV. Intuitively, there are also two solutions for the content driven FoV prediction.
\begin{itemize}
    \item \emph{Convolutional Neural Network (CNN)}. This method is to predict the viewpoint from the saliency map inferred from a CNN model. In other words, the CNN model is used to generate the saliency map from the video frame, and there is a known mapping relation between the saliency map and the FoV. For example, \cite{Lee2019NOSSDAV} proposed a tile-based approach using a saliency map that integrated the information of human visual attention with the contents.
    \item \emph{LSTM}. However, it requires much computation power on inferring the saliency map from each frame, which will also incur the playback delay. As a candidate, the LSTM model can take the past saliency maps as  the input and predict the future saliency map. The LSTM method is preferred in the current studies. For instance, \cite{Nguyen2018MM} integrated the panoramic saliency maps with user head orientation history for head movement prediction. It took saliency map and head orientation map as input and outputs the predicted head orientation of the next moment. \cite{Fan2017NOSSDAV} concurrently leveraged sensor-related and content-related features to predict the viewer fixation in the future. The sensor-related features include HMD orientations, while the content-related features include image saliency maps and motion maps.
\end{itemize}

\subsubsection*{A Hybrid User-driven and content-driven FoV Prediction}
Combining the user driven method and the content driven method, some work intended to design a user-and-content driven FoV prediction framework.
\cite{Xie2018MM_CLS} designed such a work, which integrated past fixations with cross-users' region-of-interest (ROI) found from their historical fixations. Benefiting from data-driven learning, they proposed a cross-user learning based system (CLS) to improve the precision of viewport prediction. Since users have similar ROI when watching a same video, it is possible to exploit cross-users’ ROI behavior to predict viewport. The authors used the DBSCAN clustering algorithm \cite{DBSCAN} to group together the users’ fixation that are closely packed together, i.e. fixations that have enough nearby neighbors are grouped together as a class. The viewing probability can be predicted by class.

The comparison of different FoV prediction methods is summarized in Table \ref{FoV Prediction}.

\section{Data Structure, Compression, and Projection}\label{sec_project}
A VR video is captured in every direction from a unique point, so it is essentially a spherical video.
Since current video encoders operate on a two-dimensional rectangular image, a key step of the encoding chain is to project the spherical video onto a planar surface.

\subsection{Data Structure}
Based on the view requirement, the virtual reality videos can be built on two different data structures, i.e., the 2D frame and the 3D point cloud.

\subsubsection*{2D Frame}
For 360-degree videos, the contents are generally built on 2D frames.
The 2D frames are first encoded at the content transmitter and then decoded for efficient video transmission. Afterwards, the client can project and further render the decoded 2D frames into 3D space.

\subsubsection*{3D Point Cloud}

A point cloud is a set of data points in space. The points represent a 3D shape or object. Each point has its set of X, Y and Z coordinates.
Different from conventional 2D frame structure, the volumetric video represented with 3D point cloud format enables users to watch the video with a highly immersive and interactive user experience \cite{Lee2020MobiCom}.

3D point cloud enables the ability to transmit and interact with
holographic data from remote locations across a network.
Holographic type communications (HTC)  allows remote participants to be projected as holograms into a meeting room to facilitate real-time interactions with local participants. Likewise, training and education applications can provide users the ability to dynamically interact with ultra-realistic
holographic objects for teaching purposes \cite{Clemm2020CM}.
However, HTC videos need a type of spectrum band that is currently unavailable in the millimeter-wave spectrum (5G). This situation presents significant challenges related to area or spatial spectral efficiency and the needed frequency spectrum bands for connectivity \cite{Alsharif2020Symmetry}.
Hence, a bigger radio frequency spectrum bandwidth has become necessary and can only be found at the sub-THz and THz bands, often referred to as the gap band between the microwave and optical spectra \cite{Wakunami2016NatureComm, Calvanese2019VTM}.

\subsection{Compression}
This part first introduces the titling method for the 360-degree video projected from 2D frame, and then presents the compression schemes for 3D point clouds.
We also summarize the standard for video coding and compression.

\subsubsection{Titling}
Different from standard video encoding, it is the tiling method on the projected map that should be specially considered for the VR streaming. Accordingly, the encoding scheme can be categorized into non-tiling, uniform tiling and non-uniform tiling schemes. The comparison of different encoding schemes is shown in Table \ref{Encoding Schemes}.

\subsubsection*{Non-tiling}
The non-tiling scheme encodes the VR streaming as traditional video streaming. It streams each frame as a whole to the HMD. \cite{Zhou2017MMSys} adopted the non-tiling scheme that streams the VR video as a single-tile video. Videos are split into segments temporally as in standard DASH implementations and the bitrate adaption is in the segment level. The videos are encoded by \cite{FFmpeg} with the \cite{x264} encoder.

\subsubsection*{Uniform tiling}
As only a small portion of each video frame displayed on the screen at a specific time, streaming the entire frame with the same quality is a waste of bandwidth. Many methods adopted uniform tiling schemes due to the users' dynamic viewports \cite{Feuvre2016MMSys, Qian2016AllThingsCellular, Hosseini2017VR}. To support viewport-adaptive streaming, video segments are further spatially split into equal-sized tiles. Different tiles can be allocated with different bitrate levels. Spatial representation description (SRD) has been proposed as an improvement to the MPEG-DASH standard \cite{Niamut2016MMSys}. SRD can be used to describe the spatial relationship among tiles that belong to the same temporal video segment within the DASH media presentation description file. This update in the MPEG standard promotes the deployment of tile-based streaming systems.

\subsubsection*{Non-uniform tiling}
According to \cite{Xiao2017MM_OpTile, Zhou2018INFOCOM_ClusTile, Xiao2018MM_miniView, Guan2019Pano}, fixed-size tiling approaches suffer from reduced encoding efficiency. OpTile \cite{Xiao2017MM_OpTile} is a scheme that estimates per-tile storage costs and then solves an integer linear program to obtain optimal non-uniform tiling. ClusTile \cite{Zhou2018INFOCOM_ClusTile} enhanced OpTile by clustering the collected user views. MiniView \cite{Xiao2018MM_miniView} encoded the 360-degree videos in the MiniView Layout which uses smaller-FoV projections for the purpose of saving the encoded video size while delivering similar visual qualities. Similarly, considering the perceived quality, \cite{Guan2019Pano} proposed a variable-sized tiling scheme.

\subsubsection*{Dynamic tiling}
In VR streaming, a tile could refer to data of another tile in a previous or future reference frames, leading to decoding glitches. Therefore, \cite{Zare2016MM} proposed to constrain the tiles encoding so that each tile only refers to the same tiles in previous or future frames. This reduces the complexity on the clients and in the networks.
Following this concept, \cite{Lo2017MMSys} also restricted the motion prediction when encoding the video tiles.
Similarly, \cite{Liu2019MobiCom} designed a dynamic encoding technique to dynamically determine the RoI in order to reduce the transmission latency and bandwidth consumption in the offloading pipeline.

\subsubsection{Compression}
Some work focused on the volumetric video compression represented with 3D point cloud. For example, Google Draco \footnote{Google Draco, \url{https://google.github.io/draco/}.} and Point Cloud Library (PCL) \footnote{Point Cloud Library, \url{https://github.com/PointCloudLibrary/pcl}.}, widely used open-source libraries to compress volumetric videos, decrease the data size by 4 times.

\subsubsection{Standards}
The widely used coding standards for 360-degree video are \cite{H.264}, \cite{H.265} and the extended form scalable video coding (SVC) \cite{Schwarz2007SVC}, which will be introduced as follows.

\noindent\emph{(1) H.264} is a block-oriented motion-compensation-based video compression standard. It is one of the most commonly used formats for recording, compression, and distribution of video content. Frames are encoded at the macroblock level. H.264 uses motion prediction to reduce the size of encoded data. It encodes tiles into separate video files and decodes multiple video tiles serially.

\noindent\emph{(2) H.265} is an extension of H.264. It divides a video frame into independent rectangular regions. Each region can be encoded independently. A region is essentially a tile used in tile based streaming framework. In comparison to H.264, H.265 offers  better data compression at the same level of video quality, or substantially improved video quality at the same bit rate. It decodes multiple video tiles in parallel.

\noindent\emph{(3) SVC} is an extension of H.264 and H.265. It standardizes the encoding of a high-quality video bitstream that also contains one or more subset bitstreams. SVC enables the transmission and decoding of partial bit streams to provide video services with lower temporal or spatial resolutions or reduced fidelity while retaining reconstruction quality that is high relative to the rate of the partial bit streams.

\noindent\emph{(4) PCC} is an ongoing standardization activity for point cloud compression (PCC).
The international standard body for media compression, also known as the Motion Picture Experts Group (MPEG), is planning to release in 2020 two PCC standard specifications: video-based PCC (V-CC) and geometry-based PCC (G-PCC). V-PCC and G-PCC will be part of the ISO/IEC 23090 series on the coded representation of immersive media content. The detailed description of both codec algorithms and their coding performances can be referred to \cite{graziosi_nakagami_kuma_zaghetto_suzuki_tabatabai_2020}.

\subsection{Projection}
VR videos encode information of each direction surrounding the camera location. As a result, pixels in VR videos are on the surface of a sphere. To encode the spherical pixels, the encoder maps them onto a plate surface and uses standard video codec to efficiently compress them. These mappings add redundant pixels and result in projection inefficiency. Following the work by \cite{Xiao2018MM_miniView}, the projection efficiency can be characterized as:
\begin{equation*}
\centering
    \text{Projection Efficiency}=\cfrac{\text{area on the spherical surface}}{\text{area on the calibrated projection}}.
\end{equation*}
To compute projection inefficiency, it is necessary to ensure that the size of the projection is calibrated appropriately against the unit sphere.

Universally, five projection methods are widely used for 360-degree video encoding, which include equirectangular panorama, cube map, pyramid layout, rhombic dodecahedron, and MiniView layout \cite{Corbillon2017ICC, Xiao2018MM_miniView}. For the methods except pyramid layout, it is possible to generate a viewport for any position and angle in the sphere without any information loss. However, some pixels are over-sampled, i.e., a pixel on the sphere is projected to more pixels in the projected image. Such over-sampling degrades the performance of video encoders. On the contrary, the projection into a pyramid layout causes under-sampling: some pairs of pixels on the sphere are merged into a single pixel in the projected image. This under-sampling causes distortion and information loss in some extracted viewports.
\begin{figure}
\centering
  \includegraphics[width=0.45\textwidth]{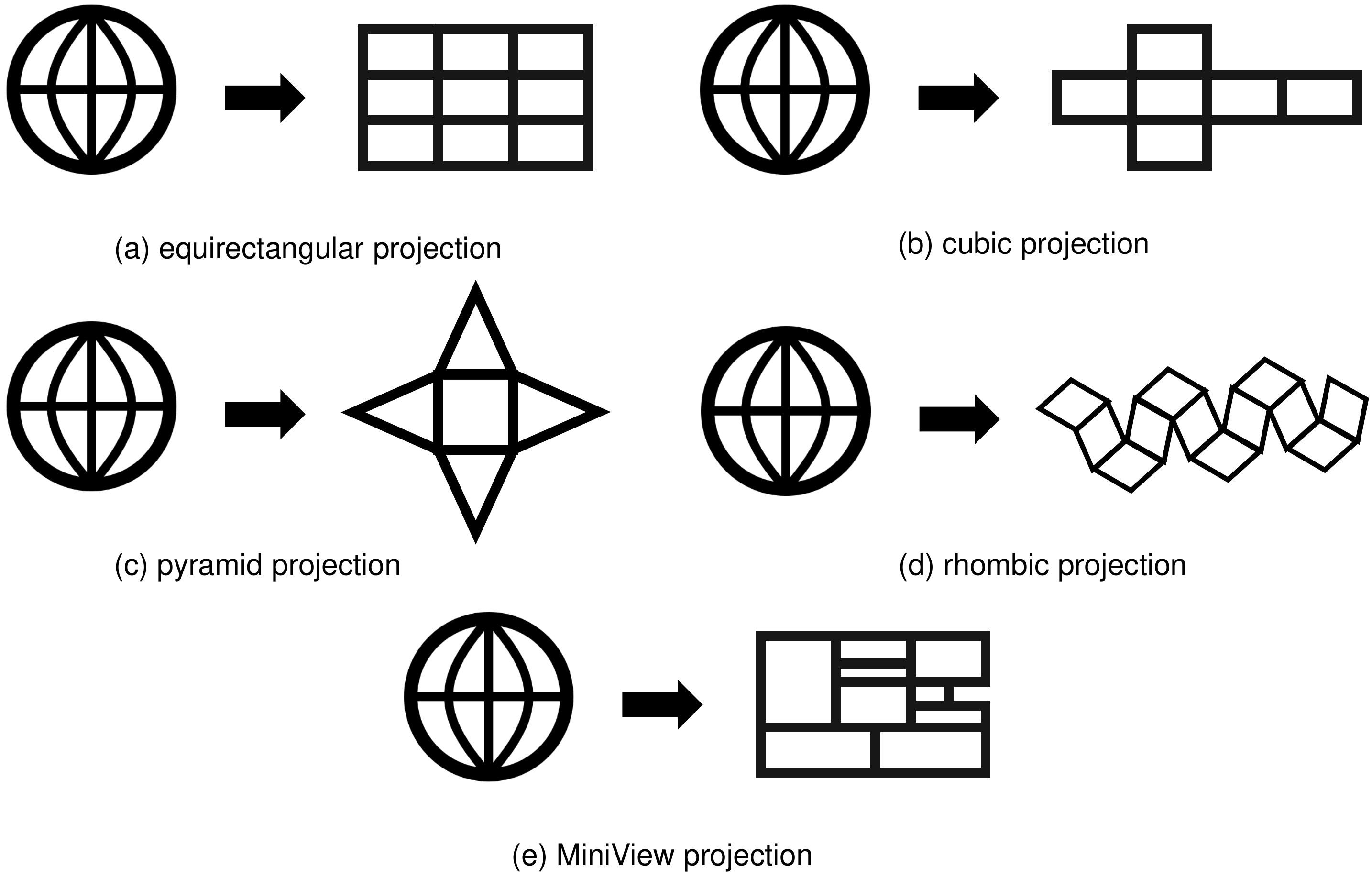}\\
  \caption{The projection methods used in existing VR systems.}\label{proj}
\end{figure}
As shown in Figure \ref{proj}, the five key projection methods are summarized as follows.

\subsubsection*{Equirectangular Projection}
As shown in Figure \ref{proj}(a), an equirectangular projection is something like the world map. The 360-degree view can be represented in a single aspect ratio image, with the expense of having distortion towards the poles.
One approach to create an equirectangular projection is to first render a sphere map, and then use a tool like \cite{ImageMagick} to convert the distortion.

\begin{table*}
\renewcommand\arraystretch{1.35}
	\caption{Encoding Schemes}\label{Encoding Schemes}
	\centering
	\resizebox{\textwidth}{!}{
		\begin{tabular}{|p{2cm}|p{2cm}|p{4cm}|c|p{4cm}|}
		\hline \bf{Category} & \bf{Literature} & \bf{Solution} & \bf{Coding} & \bf{Evaluation} \\ \hline
			\hline \multirow{1}{*}{{Non-tiling}} &  \cite{Zhou2017MMSys} & Streamed the 360-degree video as a single-tile video & H.264/FFmpeg & When the motor rotates in 5 degree increments for a total of 36 times, 46 (16.3\% of the total 282 segments) downloaded segments are wasted \\
			\hline \multirow{3}{*}{{Uniform tiling}} & Feuvre \cite{Feuvre2016MMSys} & Demonstrated a tile-based video adaptation player using the open-source \cite{GPAC} framework & HEVC & System realization without performance analysis \\
			\cline{2-5} &  \cite{Qian2016AllThingsCellular} & Proposed a cellular-friendly streaming scheme  & N/A & The bandwidth saving can reach up to 80\% when the player knows perfectly the head positions \\
			\cline{2-5} & \cite{Hosseini2017VR} & Proposed an adaptive view-aware bandwidth-efficient VR streaming framework based on the tiling features  & MPEG-DASH SRD & The approach can save bandwidth usage for up to 72\% \\
			\hline \multirow{4}{*}{{Non-uniform tiling}} & OpTile \cite{Xiao2017MM_OpTile} & Proposed a non-uniform tiling scheme by first estimating per-tile storage costs, then solving an integer linear program  & H.264/FFmpeg & Compared to the original scheme without tiling, OpTile can save 62\% to 71\% downloading volume \\
			\cline{2-5} & ClusTile \cite{Zhou2018INFOCOM_ClusTile} & Proposed a tiling approach based on clusters of collected user views & H.264/FFmpeg & ClusTile reduces 68\% to 76\% downloaded volume compared to the non-tiled scheme \\
			\cline{2-5} & MiniView \cite{Xiao2018MM_miniView} & {MiniView Layout} uses smaller-FoV projections for the purpose of saving the encoded video size while delivering similar visual qualities & H.264/ffmpeg360 & MiniView saves storage space with the median and mean ratios being 0.84 and 0.83 \\
			\cline{2-5} & Pano \cite{Guan2019Pano} & Proposed a variable-sized tiling scheme in order to strike a balance between the perceived quality and video encoding efficiency & H.264/FFmpeg & Pano achieves the same PSPNR with 41-46\% less bandwidth consumption than the viewport-driven baseline \\
			\hline
	\end{tabular}}
\end{table*}

\subsubsection*{Cubic Projection}
As shown in Figure \ref{proj}(b), cubic is a type of projection for mapping a portion of the surface of a sphere (or the whole sphere) to flat images. The images are arranged like the faces of a cube, and this cube is viewed from its center.
Four cube faces cover front, right, back and left, one the zenith and one the nadir, each of them having $90^\circ\times 90^\circ$ FoV. In each cube face, all straight lines stay straight, hence it is quite suitable for editing.
To address the projection inefficiency of cubic projections,  \cite{EAC} recently proposed equi-angular cube (EAC). EAC attempts to address the problem of pixel inefficiency in cube faces by distorting the standard cubic projection to reduce projection inefficiency.

\subsubsection*{Pyramid Projection}
As shown in Figure \ref{proj}(c), in the pyramid projection, the spherical video is projected onto a pyramid layout from central points to generate a set of video representations. The base of pyramid is in the viewing direction with full resolution, while the rest pixels are projected onto the sides of pyramid with decreasing resolution. However, the pyramid projection is especially sensitive to head movements, because sides of a pyramid corresponding to non-viewing directions are rendered with low resolution. Besides, since there are multiple viewports rendered and stored for the video, pyramid projection introduces extra overheads to a certain extent.

\subsubsection*{Rhombic Projection}
As shown in Figure \ref{proj}(d), a rhombic projection can be dissected with its center into 4 trigonal trapezohedra \cite{Fu2009TMM}. These rhombohedra are the cells of a trigonal trapezohedral honeycomb. This is analogous to the dissection of a regular hexagon dissected into rhombi, and tiled in the plane as a rhombille.

\subsubsection*{MiniView Layout}
Recently, \cite{Xiao2018MM_miniView} proposed the MiniView layout, which improves bandwidth efficiency of VR streaming by reducing the quantity of over-represented pixels in the spherical-to-2D projection. As shown in Figure \ref{proj}(e), MiniView encodes a small portion of the sphere by applying the rectilinear projection with a small FoV.

\section{Adaptive VR Streaming}\label{sec_stream}
In this section, we first summarize the state-of-the-art bitrate adaption solutions for VR systems. Then we enumerate the focused QoE metrics for VR applications.

\begin{table*}
\renewcommand\arraystretch{1.35}
	\caption{A summary of VR Streaming Schemes}\label{Adaptive Bitrate Schemes}
	\centering
	\resizebox{\textwidth}{!}{
		\begin{tabular}{|p{2cm}|p{2cm}|p{3cm}|p{3.5cm}|p{3cm}|p{3cm}|p{3cm}|}
			\hline \bf{Category} & \bf{Literature} & \bf{Solution} & \bf{Testbed} & \bf{Computational Requirements} & \bf{Latency} & \bf{QoE} \\
			\hline \hline \multirow{3}{*}{\shortstack{Heuristic-based \\Bitrate Adaptation}} & \cite{Graf2017MMSys}  & Explore various options about full delivery and partial delivery over HTTP & libVR, tileMuxer, tileTools, Android-based Tiled Player, Web-based Tiled Player & Mobile phone, desktop & No comparison was made & The average V-PSNR is about 30dB to 40dB.  \\
			\cline{2-7} & \cite{Petrangeli2017MM} & The tiles are divided into different groups. The tiles of viewport are streamed at a higher quality & Samsung Galaxy S7, Gear VR, MP4Box & Android devices & The total freeze time is less than 3 seconds when watching 60 seconds video & No comparison was made \\
			\cline{2-7} & \cite{CostaFilho2018MMSys} & The tiles are divided into different zones. The tiles of viewport are streamed at a higher quality & Linux Ubuntu 14.04 operating system, web server (Apache 2 2.4.18-2) & a quad-core E3-1220v3 (3.1GHz) processor with 16GB RAM & The average normalize stall time is about 90ms & The residual error of QoE estimation is smaller than 0.03922 for over 90\% of the cases \\
			\hline \multirow{3}{*}{\shortstack{Optimization-based\\Bitrate Adaptation}} &  Rubiks \cite{He2018MobiSys_Rubiks} & Use an optimization framework to optimize the user QoE in chunk level & Android APP, Samsung Galaxy S7, Huawei Mate9, a laptop equipped with a wireless NIC Intel AC-8260 & Android phone, laptop & The average rebuffering time of Rubiks is 1.2s  & Rubiks outperforms YouTube, FoV-only and FoV+ by 14\%, 69\%, and 26\% respectively \\
			\cline{2-7} & CLS \cite{Xie2018MM_CLS} & Construct a Multiple-Choice Knapsack model to determine the rates for tiles & Simulation & Parameter assumption & No comparison was made & The average V-PSNR is about 25dB to 32dB  \\
			\cline{2-7} & Pano \cite{Guan2019Pano} & Optimize the bitrate of each chunk by maximizing the overall perceived quality with total size constraint & An Oculus headset (Qualcomm Snapdragon 821 CPU, 3GB RAM, Adreno 530 GPU) as the client, a Windows Server 2016-OS desktop & an Intel Xeon E5-2620v4 CPU, 32GB RAM, Quadro M2000 GPU  & No comparison was made & The PSPNR value is about 60 to 70 dB \\
			\cline{2-7} & Flare \cite{Qian2018Flare}  & Optimize the quality level for each to-be-fetched tile by jointly considering different QoE objectives & Android smartphones, Ubuntu 14.04 and 16.04 & SGS7, SGS8, Samsung Galaxy J7, desktop & No comparison was made & Flare yields up to 18× quality level improvement on WiFi video quality enhancement (up to 4.9) on LTE \\
			\hline \multirow{5}{*}{\shortstack{DRL-based\\Bitrate Adaptation}} & Plato \cite{Jiang2018LCN}  & Use A3C-based model to determine the bitrates for both viewport and non-viewport & Simulation & Parameter assumption & The average rebuffering time is about 3.7 in the defined QoE & The average value of defined QoE is about -1.4 \\
			\cline{2-7} & DRL360 \cite{Zhang2019INFOCOM_DRL360}  & The A3C-based model adaptively allocates rates for the tiles of the future video frames & Linux Ubuntu 14.04 operating system, web server (Apache 2 2.4.18-2) & a quad-core E3-1220v3 (3.1GHz) processor, with 16GB of RAM & No comparison was made & DRL360 outperforms other methods by 20\%-30\% \\
			\cline{2-7} & \cite{Xiao2019ACM_TURC}  & Use A3C-based model to determine the quality level of the tiles in estimated FoV & Simulation & Android devices, desktop & The average rebuffering time is reduced by 0.002s while playing each video segment of 1s & The average value of defined QoE is about 1.15 and the average QoE is 12\% higher than the others \\
			\cline{2-7} & \cite{Kan2019ICASSP}  & Use A3C-based model to determine the bitrates for each estimated FoVs and the tiles in the same FoV are allocated with the same bitrate & Simulation & Parameter assumption & The average value of rebuffering time is about 1.0 in the defined QoE & The average value of defined QoE is 8.3 and is at least 13.7\% higher than other solutions \\
			\cline{2-7} & SR360 \cite{Chen2020NOSSDAV_SR360} & Use A3C model to determine the tile bitrates and enhance the video with SR & Simulation & An Nvidia GTX 1080 GPU & The average value of rebuffering time is about 0.42 in the defined QoE  & The average  QoE is about 7.1 and SR360 outperforms others by at least 30\% \\
			\hline
	\end{tabular}}
\end{table*}

\subsection{Bitrate Adaption Design}
The adaptive VR streaming strategy can be broadly classified into three categories: 1) heuristic-based bitrate adaptation, 2) optimization theory-based bitrate adaptation and 3) deep reinforcement learning (DRL)-based bitrate adaptation \cite{Kua2017CST}. We will look at examples of all three bitrate adaptation approaches. The comparison of different streaming schemes is shown in Table \ref{Adaptive Bitrate Schemes}.

\subsubsection{Heuristic-based Bitrate Adaptation}
The heuristic bitrate adaption schemes take maximizing bandwidth utilization as an objective and allocate the highest quality to the tiles in viewport \cite{Graf2017MMSys, Petrangeli2017MM, CostaFilho2018MMSys}.
\cite{Petrangeli2017MM} divided the tiles into different groups. With the constrained bandwidth, the tiles of viewport were streamed at a higher quality and the others were streamed at a lower quality.
\cite{Graf2017MMSys} explored various options about full delivery and partial delivery to enable the bandwidth efficient adaptive streaming of omnidirectional video over HTTP.

\subsubsection{Optimization Theory-based Bitrate Adaptation}
This category solves the bitrate adaption problem as an optimization problem. Rubiks \cite{He2018MobiSys_Rubiks} used an MPC-based optimization framework to optimize the user QoE in chunk level. CLS \cite{Xie2018MM_CLS} constructed a Multiple-Choice Knapsack model to determine the rates for tiles. Pano \cite{Guan2019Pano} determined the bitrate of each chunk and maximized the overall perceived quality with total size constraint based on MPC. Flare \cite{Qian2018Flare} determined the quality level for each to-be-fetched tile by jointly considering different QoE objectives.

\begin{figure}[h]
\centering
  \includegraphics[width=0.5\textwidth]{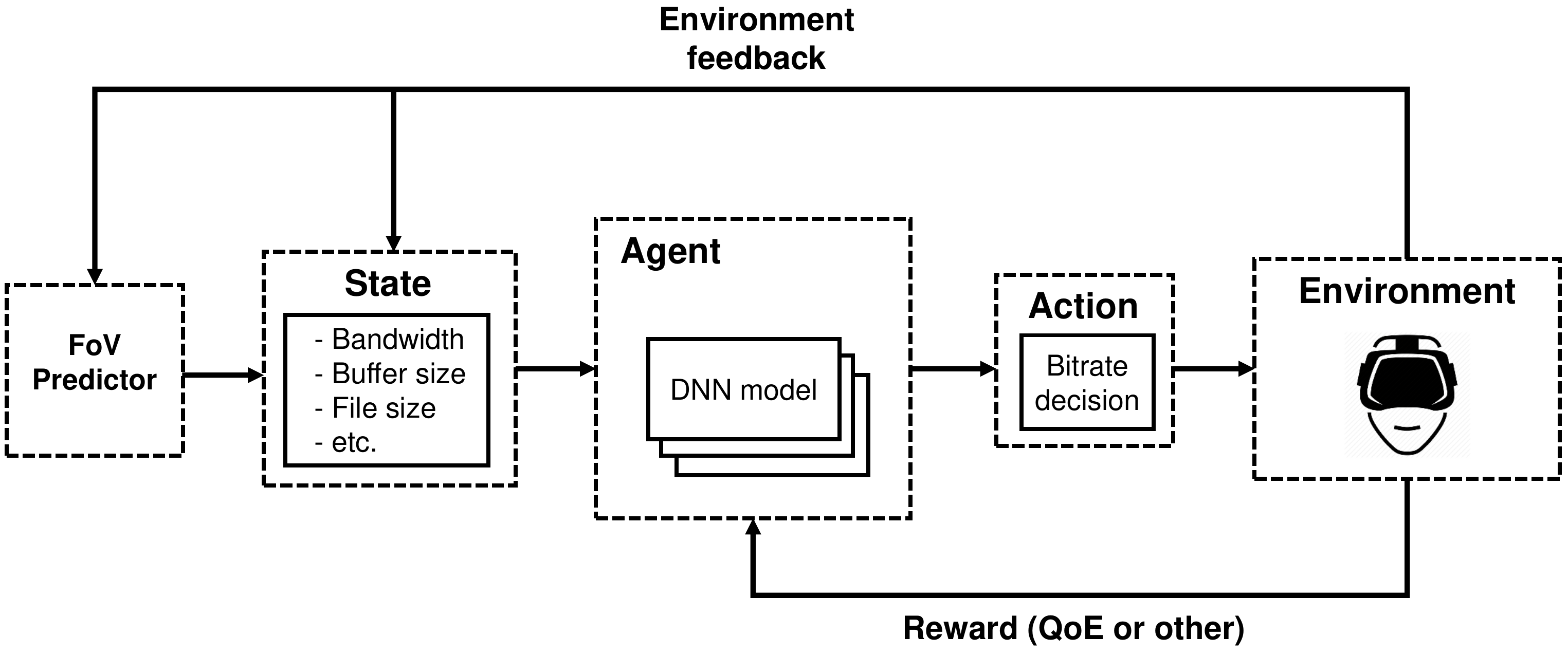}\\
  \caption{DRL-based bitrate adaptation for VR streaming \cite{Zhang2019INFOCOM_DRL360}.}
  \label{DRLframework}
\end{figure}

\subsubsection{DRL-based Bitrate Adaptation}
The deep reinforcement learning achieves great success in traditional video streaming. It also performs prominent at VR streaming in this two years \cite{Jiang2018LCN, Zhang2019INFOCOM_DRL360, Xiao2019ACM_TURC, Kan2019ICASSP, Chen2020NOSSDAV_SR360}. These schemes directly optimized the use QoE and they all adopt the A3C to train the model.
\cite{Jiang2018LCN} used A3C-based model to determine the bitrates for both viewport and non-viewport areas. DRL360 \cite{Zhang2019INFOCOM_DRL360} adaptively allocated rates for the tiles of the future video frames based on the observations collected by client video players.
\cite{Xiao2019ACM_TURC} determined the quality level of the tiles in estimated FoV.
\cite{Kan2019ICASSP} determined the bitrates for multiple estimated FoVs. SR360 \cite{Chen2020NOSSDAV_SR360} adaptively determined the tile bitrate inside the viewport and enhanced the tile quality by super-resolution (SR) at client side.

\subsection{QoE Metrics}
\begin{figure}
\centering
  \includegraphics[width=0.45\textwidth]{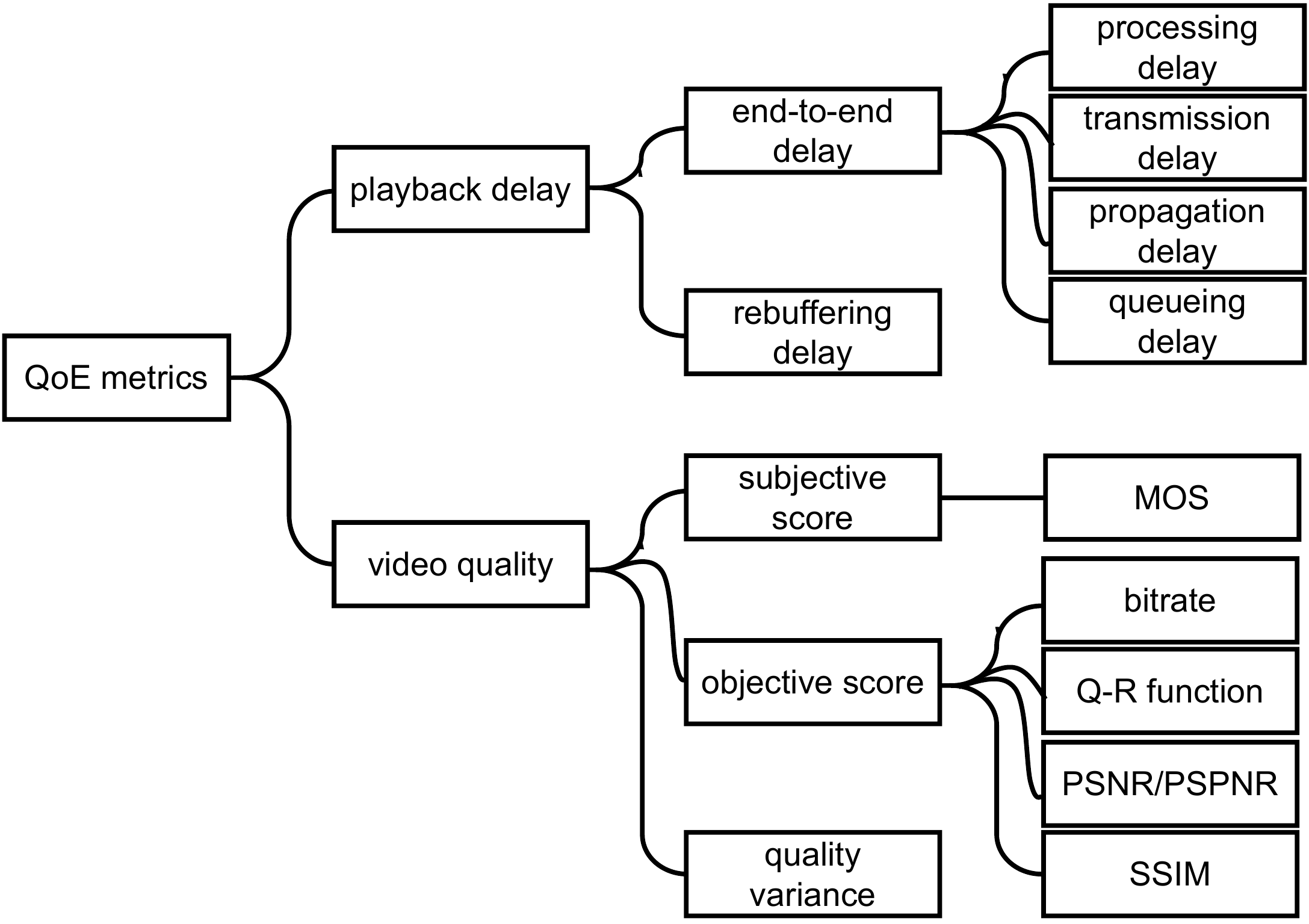}\\
  \caption{QoE metrics for VR streaming.}
  \label{QoE}
\end{figure}

In this section, we discuss QoE metrics for VR streaming in two key respects: playback delay and video quality. Playback delay can be divided into the delay caused by video transmission and processing. The transmission delay is mainly affected by packet propagation, while the processing delay can be further classified into coding delay and rebufferring delay. Video quality can be categorized into three types: subjective score, objective score and the variance on video quality. Figure \ref{QoE} shows QoE metrics defined above. In the following we describe each of these QoE metrics with the review and discussion of important works in Table \ref{QoE_table}.

\subsubsection{Playback Delay}
Following \cite{Baldi2000ToN, Bovy2002PAM, Erbad2010MMSys, Kumar2017RTSS}, we divide the playback delay into two key components, i.e., the end-to-end delay and the rebuffering delay, which will be introduced as follows.

\subsubsection*{End-to-end Delay}
In VR systems, the end-to-end delay mainly includes processing delay, transmission delay, propagation delay, queueing delay, and other kinds of delay. The propagation delay is the time for a packet to reach its destination, e.g., HMD. The rebuffering delay is the additional time needed for packet retransmission. These five key kinds of network delays are introduced as follows.

\noindent\emph{(1) Processing Delay}.

The processing delay for VR streaming transmission can be represented by coding/decoding delay.
Current streaming strategies approaches found that traditional decoding and existing tile-based decoding cannot support 8K or higher video resolution.
Correspondingly, \cite{He2018MobiSys_Rubiks} suggested that we do not have the luxury to allocate a tile to each decoding thread, but should have a different task-to-thread assignment.

\noindent\emph{(2) Transmission Delay}.

The transmission delay is the time needed to transmit an entire packet, from the first bit to the last bit. The transmission delay is determined primarily by the number of bits and the capacity of the transmitter device.

\noindent\emph{(3) Propagation Delay}.

The propagation delay is the time to propagate a bit through the communication link.
It can be computed as the ratio between the link length and the propagation speed over the specific medium.
Generally, the propagation delay is equal to $d/v$ where $d$ is the distance between the transmitter and the receiver and $v$ is the wave propagation speed.

\noindent\emph{(4) Queueing Delay}.

The queueing delay is defined as the waiting time of packets in the buffer until they can be executed. Queueing delay may be caused by delays at the originating switch, intermediate switches, or the receiver servicing switch.

\noindent\emph{(5) Other Sources of Delay}.

There are other sources of delay:
\emph{i)}. Packetization Delay. Packetization delay is the time taken to fill a packet payload with encoded data. Packetization delay can also be called accumulation delay, as the data samples accumulate in a buffer before they are released.
\emph{ii)}. Serialization Delay. Serialization delay is the fixed delay required to clock a  data frame onto the network interface. It is directly related to the clock rate on the trunk. At low clock speeds and small frame sizes, the extra flag needed to separate frames is significant.

\subsubsection*{Rebuffering Delay}
The playback and downloading strategy should adapt to the buffer occupancy. If the buffer occupancy reaches 0 (i.e., no chunk is left in the buffer), the playback will suffer from \emph{rebuffering} \cite{Zhang2019INFOCOM_DRL360}. The playback will continue until there is at least one chunk in the buffer.
Let $b_k$ denote the buffer occupancy when the $k$-th chunk has been downloaded completely. Therefore, after the startup stage, the buffer occupancy changes according to the downloading of the new chunk and consumption of chunks in the buffer, i.e.,
\begin{equation}
    b_k=b_{k-1}-\cfrac{R_k T}{C_k}+T
\end{equation}
where $T$ is the length of a video chunk, $R_k$ is the sum bitrates for caching chunk $k$, and $C_k$ is the downlink rate for chunk $k$ that can be estimated \cite{Zhang2019INFOCOM_DRL360}.

Following \cite{Xie2017MM_360ProbDASH}, we prefer to let the buffer occupancy stay at least $B_{\text{target}}$, that is $b_k\geq B_{\text{target}}$. Then the sum bitrates for caching video chunk $k$ should satisfy:
\begin{equation}
    R_k\leq\cfrac{C_k}{T}(b_{k-1}-B_{\text{target}}+T).
\end{equation}
Generally, we set a lower bound $R_{\min}$ to $R_k$, then we have:
\begin{equation}\label{Rlimit}
    R_k=\max\left\{R_{\min},\cfrac{C_k}{T}(b_{k-1}-B_{\text{target}}+T)\right\}.
\end{equation}

\subsubsection{Video Quality}
The term ``video quality" is defined in terms of fidelity, i.e., how closely a processed or delivered signal matches the original source (or reference) signal. The main concern is to detect and quantify any distortions that have been introduced into the signal as it passes through a network or device. The video quality metric can be further split into three parts, i.e., subjective score, objective score, and quality variance.

\subsubsection*{Subjective Quality}
Subjective testing is the only proven way to evaluate video quality \cite{Clarity2014WhitePaper}. For example, human subjects are asked to assess the overall quality of the processed sequence with respect to the original (reference) sequence. Unfortunately, this mode of testing is very expensive, time-consuming, and often impractical. The subjects score the processed video sequence on a scale corresponding to their mental measure of the quality, termed \emph{Mean Opinion Score (MOS)}.
As shown in Table \ref{MOS_table}, when the MOS score is on a 1 to 5 scale, corresponding opinions for these scores are Unacceptable, Poo, Fair, Good and Excellent, respectively.
Note that the MOS values are different for various stalling events under different bitrate levels \cite{ANWAR2020SCIENCECHINA}.
\cite{Schatz2017QoEMX} showed that video stall severely impacts the QoE, much stronger than other typical impairments like initial delay or quality switching.
Furthermore,  high-quality 360 videos at 4K
resolutions can provide good perceptual quality score
(around 4.5 MOS) to users \cite{Tran2017MMSP}.

Specifically, as one of the key components of MOS, cybersickness is the feeling of nausea and dizziness when watching virtual reality video.
\cite{Anwar2020Access} subjectively evaluated the impact
of three QoE-affecting factors (gender, user's interest, and user's familiarity with VR) on cybersickness. The statistical results from the subjective experiment provided various findings on QoE impact factors in terms of perceptual quality and cybersickness.
To predict the QoE, a neural network model was trained and tested on cybersickness data obtained from the subjective test.

\begin{table}[h]
\caption{Human subjects versus MOS}\label{MOS_table}
\centering
\linespread{1}\selectfont
\begin{tabular}{cccccc}
\hline MOS & 1 & 2 & 3 & 4 & 5 \\
\hline Opinion & Unacceptable & Poor & Fair & Good & Excellent \\
\hline
\end{tabular}
\end{table}

\begin{table*}
\renewcommand\arraystretch{1.35}
\caption{QoE metrics used in existing VR systems}\label{QoE_table}
\centering
\linespread{1}\selectfont
\begin{tabular}{|c|c|c|c|c|c|}
\hline \bf{Literature} & \bf{Rebuffer Delay} & \bf{Coding Delay} & \bf{Subjective} & \bf{Video Quality} & \bf{Quality Variance} \\
\hline\hline Optile \cite{Xiao2017MM_OpTile}  & $\times$ & $\times$ & $\times$ & bitrate & $\times$ \\
\hline ClusTile \cite{Zhou2018INFOCOM_ClusTile}  & $\times$ & $\times$ & $\times$ & bitrate & $\times$ \\
\hline \cite{Corbillon2017MM}  & $\times$ & $\times$ & $\times$ & bitrate & $\times$ \\
\hline BAS360 \cite{Xiao2018INFOCOM}  & $\times$ & $\times$ & $\times$ & bitrate & $\times$ \\
\hline \cite{Xiao2018MM_miniView}  & $\times$ & $\times$ & $\times$ & bitrate & $\times$ \\
\hline CLS \cite{Xie2018MM_CLS}  & $\times$ & $\times$ & $\times$ & bitrate & $\times$ \\
\hline \cite{Xu2019TPAMI}  & $\times$ & $\times$ & $\times$ & bitrate & $\times$ \\
\hline \cite{Yi2019NOSSDAV}  & $\times$ & $\times$ & $\times$ & bitrate & $\times$ \\
\hline Falre \cite{Qian2018Flare}  & $\times$ & $\times$ & $\times$ & bitrate & intra-/inter-chunk \\
\hline 360ProbDASH \cite{Xie2017MM_360ProbDASH}  & $\times$ & $\times$ & $\times$ & bitrate & temporal/spatial \\
\hline Plato \cite{Jiang2018LCN}  & \checkmark & $\times$ & $\times$ & bitrate & temporal/spatial \\
\hline \cite{Sun2018MMSys} & \checkmark & $\times$ & $\times$ & Q-R function & $\times$ \\
\hline \cite{Liu2018MobiSys}  & \checkmark & \checkmark & $\times$ & bitrate & $\times$ \\
\hline Rubiks \cite{He2018MobiSys_Rubiks} & \checkmark & \checkmark & $\times$ & SSIM & temporal \\
\hline DRL360 \cite{Zhang2019INFOCOM_DRL360} & \checkmark & \checkmark & $\times$ & SSIM & temporal \\
\hline Pano \cite{Guan2019Pano}  & \checkmark & \checkmark & $\times$ & PSPNR & JND \\
\hline \cite{VandenBroeck2017MM} & $\times$ & $\times$ & \checkmark & SUS score & $\times$ \\
\hline \cite{Li2018MM} & $\times$ & $\times$ & MOS/DMOS & PSNR/SSIM & temporal \\
\hline
\end{tabular}
\end{table*}

\subsubsection*{Objective Quality}
As the subjects are difficult to obtain, a number of algorithms have been developed to estimate video quality by using mathematical analysis in place of human observers.
The widely used metrics include sum bitrate, quality-rate (Q-R) function, peak signal-to-noise ratio (PSNR), and structural similarity (SSIM).

\noindent\emph{(1) Sum Bitrate}

In Table \ref{QoE_table}, we can observe that the bitrate metric is almost used in most literature, except for work focusing on subjective scores (e.g., \cite{VandenBroeck2017MM, Li2018MM}).
The sum bitrate is the metric that is most focused in the schedule of VR streaming.

In the VR streaming problem formulation, the bitrate metric, as the target scheduling parameter, can be regarded as the constraint or the objective. In \cite{Corbillon2017MM}, the scheduling constraint is the sum of the byte-stream sizes for every area is equal to $B$, where $B$ denotes the transmission bandwidth. Generally, the sum bitrates allocated to tiles is less than the transmission bandwidth \cite{Jiang2018LCN}. In \cite{He2018MobiSys_Rubiks}, the objective of the VR streaming design is to maximize the bandwidth saving, i.e., minimize the bandwidth cost.

\noindent\emph{(2) Quality-Rate (Q-R) Function}

Instead of directly using the sum bitrate metric, some work defined the VR streaming quality as a mapping function from the bitrate. For example, according to \cite{Xue2014ICMEW, Duanmu2017JSTSP},  user QoE has a logarithmic relation following the rate distortion theory, i.e.,
\begin{equation}
    \text{QoE}_{i}(r_i^{(k)})=\log(1+r_i^{(k)}/\eta)
\end{equation}
where $r_i^{(k)}$ represents the allocated bitrate for tile $i$ of chunk $k$ and $\eta$ is a constant obtained from field measurements. In total, we have the QoE on video quality for all tiles as:
\begin{equation}
    \text{QoE}(r_1^{(k)},r_2^{(k)},\cdots,r_N^{(k)}) = {\sum_{i=1}^{N} \delta_i^{(k)}\log(1+r_i^{(k)}/\eta)}.
\end{equation}

\noindent\emph{(3) Peak Signal-to-Noise Ratio (PSNR)}

Peak signal-to-noise ratio, often abbreviated PSNR, is an engineering term for the ratio between the maximum possible power of a signal and the power of corrupting noise that affects the fidelity of its representation. Because many signals have a very wide dynamic range, PSNR is usually expressed in terms of the logarithmic decibel scale.
PSNR is most easily defined via the mean squared error (MSE). Given a noise-free $m\times n$ monochrome image $I$ and its noisy approximation $K$, MSE is defined as:
\begin{equation}
    \text{MSE}=\cfrac{1}{mn}\sum_{i=0}^{m-1}\sum_{j=0}^{n-1}[I(i,j)-K(i,j)]^2
\end{equation}
The PSNR (in dB) is defined as:
\begin{equation}
    \text{PSNR}=20\log_{10}(\text{MAX}_I)-10\log_{10}(\text{MSE})
\end{equation}
where $\text{MAX}_I$ is the maximum possible pixel value of the image. When the pixels are represented using 8 bits per sample, this is 255. More generally, when samples are represented using linear PCM with $B$ bits per sample, $\text{MAX}_I$ is $2^B-1$.
A PSNR value of 35dB is generally considered good.

Recently, \cite{Yu2015ISMAR} proposed an improved PSNR metric, named viewport PSNR (V-PSNR), which can directly indicate the quality of content in the user’s viewport.

\noindent\emph{(4) Structural Similarity (SSIM)}

The SSIM index is a method for predicting the perceived quality of all kinds of digital images and videos. The SSIM index is calculated on various windows of an image. The measure between two windows $x$ and $y$ of common size $N\times N$ can be calculated as:
\begin{equation}
    \text{SSIM}(x,y)=\cfrac{(2\mu_x\mu_y+c_1)(2\sigma_{xy}+c_2)}{(\mu_x^2+\mu_y^2+c_1)(\sigma_x^2+\sigma_y^2+c_2)}
\end{equation}
where $\mu_x$ and $\sigma_x^2$ are the average and the variance of $x$, and $\sigma_{xy}$ is the covariance of $x$ and $y$. Two variables $c_1=(k_1L)^2$ and $c_2=(k_2L)^2$ are set to stabilize the division with weak denominator, where $L$ is the dynamic range of the pixel-values and $k_1=0.01$, $k_2=0.03$ following \cite{Wang2003SSIM}.

The SSIM formula is based on three comparison measurements between the samples of $x$ and $y$: luminance, contrast and structure. The SSIM actually measures the perceptual difference between two similar images. It cannot judge which of the two is better: that must be inferred from knowing which is the ``original" and which has been subjected to additional processing such as data compression. Unlike PSNR, SSIM is based on visible structures in the image.

However, the SSIM index achieves the best performance when applied at an appropriate scale (i.e. viewer distance or screen height). Calibrating the parameters, such as viewing distance and picture resolution, create the most challenges of this approach. To rectify this, multi-scale, structure similarity (MS-SSIM) has also been defined \cite{Wang2003SSIM}. In MS-SSIM, the picture is evaluated at various resolutions and the result is an average of these calibrated steps. It has been shown that MS-SSIM outperforms simple SSIM even when the SSIM is correctly calibrated to the environment and dataset.

\subsubsection*{Quality Variance}
If the quality of the content is not smooth, it will decrease the users' QoE, and we calculate this value according to the coefficient of variation of quality of content in the viewport.
It is also essential to consider the measure of quality switching (i.e., the temporal variation) between consecutive chunks. The rate changes between two consecutive chunks may impulse users with physiological symptoms such as dizziness and headache. Hence, quality switch should not be drastic, which can be measured by:
\begin{equation}
    \text{QoE}(r_1^{(k)},r_2^{(k)},\cdots,r_N^{(k)})=\sum_{i=1}^{N}\left(r_i^{(k)}-r_i^{(k-1)}\right)^2
\end{equation}
where $r_i^{(k)}$ denotes rate of tile $i$ at the $k$-th chunk.

Peak signal-to-perceptible-noise ratio (PSPNR) is an extension of PSNR \cite{Chou1995PSPNR}. It improves the classic PSNR by filtering out quality distortions that are imperceptible by users. PSPNR is proposed to assess the quality of the compressed image by taking the perceptible part of the distortion into account. The key to PSPNR is notion of just-noticeable-difference (JND) \cite{Zhao2011TCSVT}, which is defined by the minimal changes in pixel values that can be perceived by viewers. The perceptible distortion of the compressed image is referred to the part by which the distortion is noticeable, or the part of distortion that exceeds the threshold given by the JND profile.

\section{VR Applications in IoT}\label{sec_app}

IoT is a global infrastructure, which enables advanced services by interconnecting physical and virtual things based on evolving inter-operable information and communication technologies \cite{sector2012series}.
Nowadays, IoT applications appear in all aspects of our daily life, including entertainment, smart education, smart healthcare, smart home, smart retail, smart city, smart grids, and industrial Internet. As a complementary technology, VR can help improve the user experience for these IoT applications, where we provide some examples as follows.

\subsection{VR for Immersive Entertainment}
With VR technology, users can obtain fairly immersive quality of experience on video watching and interactive gaming. Most of existing VR systems focused on providing this kind of immersive entertainment experience for IoT applications. Generally, the resolution of VR video should be 8K or even higher with a bitrate more than 100Mbps to provide fairly good experience \cite{Song2020TMM}. Streaming VR video at such resolution is usually not a trivial task, especially through the Internet.
How to deliver the VR video with limited bandwidth is the most critical issue to be solved for guaranteeing a real immersive entertainment experience.

Existing applications mainly investigate on-demand VR video streaming from content providers, while the area of interactive VR video telephony remains mostly unexplored.
\cite{Xie2017CoNEXT_POI360} designed POI360, a portable interactive VR video telephony system that jointly investigates both panoramic video compression and responsive video stream rate control.

\subsection{VR for Smart Education}
The demand for VR education is growing since traditional lectures have problems in motivating and engaging students.
\cite{Slavova2018VR} conducted a comparative study on students' performance in a standardised assessment when course content is delivered using VR and conventional lecture slides. Results show that improvement in social interaction and productivity tools in VR are essential for its greater impact in higher education.
\cite{Kim2019VR} introduced the first VR MOOC learning management system.
\cite{Wei2019VR} designed a VR course with a teamwork project assignment for undergraduates, providing them with smart learning experiences.

VR also has the potential to improve surgical planning.
\cite{Fiederer2019VR} presented a framework for VR neurosurgery, and showcased the potential of such a framework using clinical data of two patients and research data of one subject.
\cite{Todsen2018VR} developed a VR video to demonstrate how to use the surgical safety checklist in the operating room. With the use of VR technology, VR videos give the medical students a realistic experience of the operating room where they can observe the teamwork performed to ensure patient safety during surgery.
With researchers and companies finding new applications of VR in the medical arena, insurance companies are now becoming interested in adding VR to the health care arsenal \cite{Mertz2019Pulse}.
In addition, \cite{Banks2004JNCA} undertook the VR technology into simulating both visual and auditory hallucinations.

\subsection{VR for Smart Home}
\begin{figure}
\centering
  \includegraphics[width=0.4\textwidth]{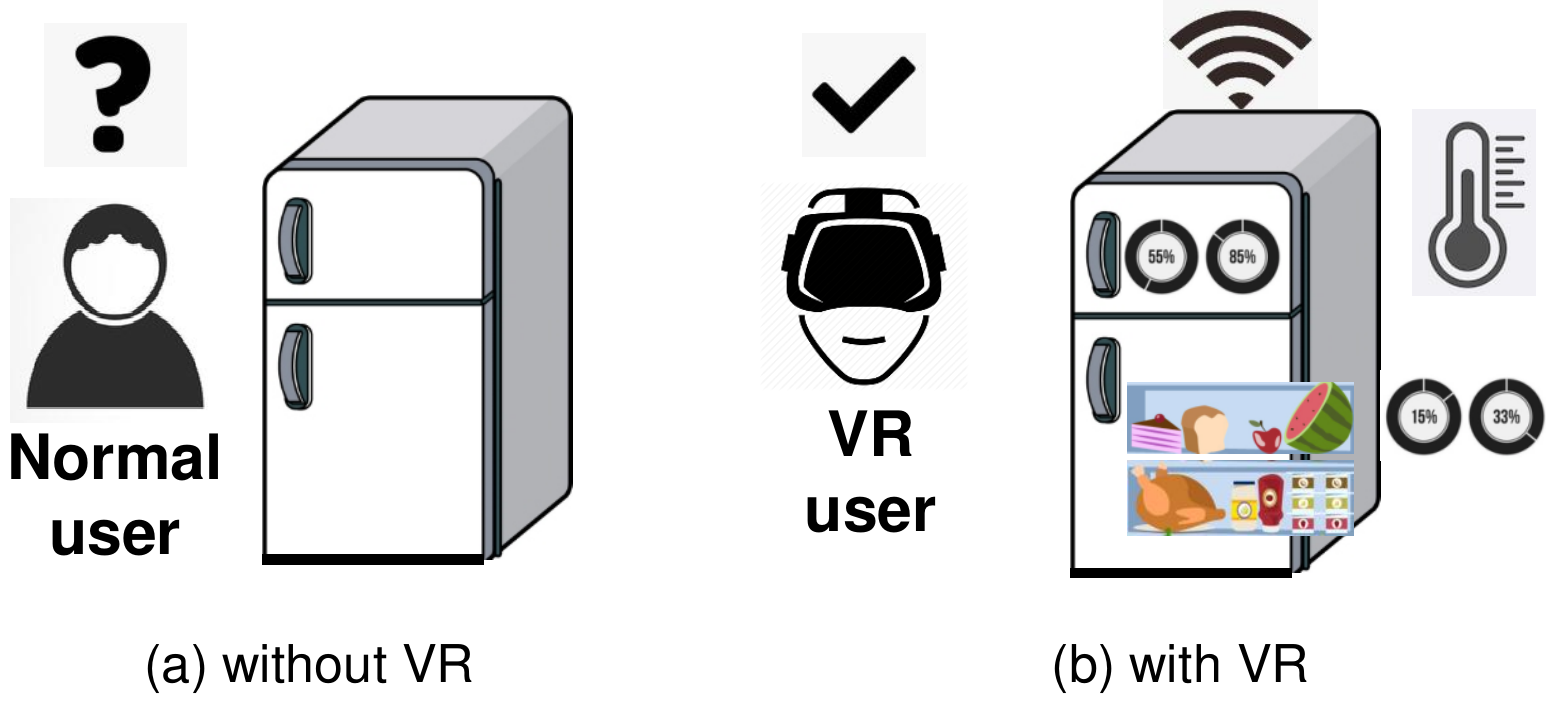}\\
  \caption{The smart fridge without and with virtual reality.}\label{fridge}
\end{figure}

Smart home is regarded as one of the most popular IoT applications.
Tremendous household appliances (e.g., cameras, clocks, lights, doorbells, windows, and heaters) can be connected to the Internet, forming a smart home.
These appliances can generate sensing information, and users can control everything at home.
However, it is still rather simple for the communication between the house owner and smart appliances, where users often need to check the status of appliances by accessing a specific website.
VR provides an efficient visible method to display the sensing data of smart appliances.
For example, Figure \ref{fridge} presents a smart fridge without (or with) virtual reality.
Without VR, the user can only access the data sensed from the fridge on a website with a specific user account.
On the contrary, with VR, the data can be directly presented in front of users, e.g., stored vegetables, temperature, and fridge space utilization.

\subsection{VR for Smart City}
Internet of Vehicle (IoV) helps a vehicle connect to nearby vehicles or traffic infrastructures using a wireless network. It helps the driver locate his (or her) vehicle as well as other vehicles in the nearby region. With IoV, a vehicle can detect the speed and distance of nearby vehicles and adjust its speed accordingly. The global IoV market is projected to reach \$208,107 million by 2024, with a growth rate of 18\% \cite{IoVMarket}.

\begin{figure}
\centering
  \includegraphics[width=0.4\textwidth]{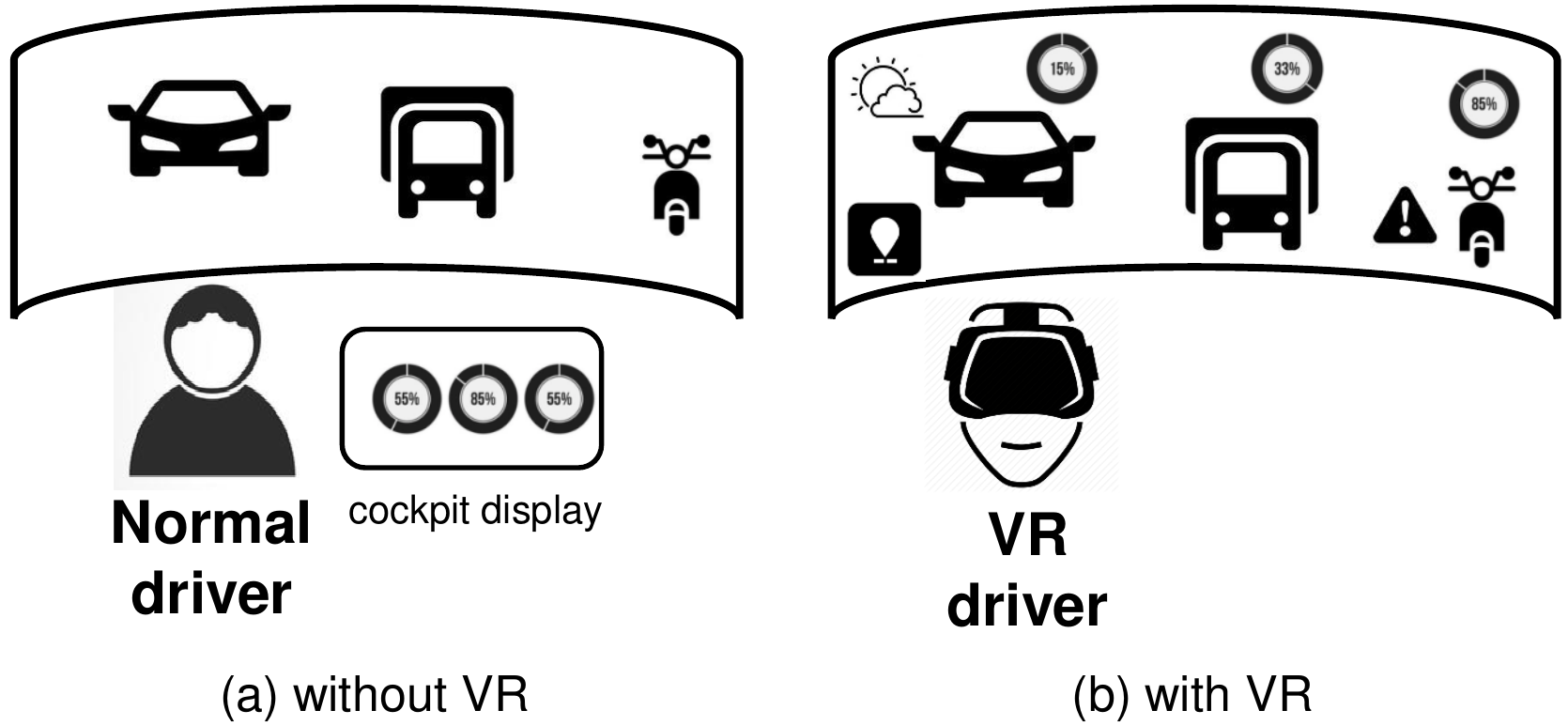}\\
  \caption{VR for the Internet of vehicles.}\label{IoV}
\end{figure}

Current IoV applications provide traffic safety service by showing warning messages on the cockpit display, which might distract the driver's attention.
Through virtual presentation of the sensed data, VR helps drivers obtain visible information of the conditions of nearby vehicles and some additional information.
As shown in Figure \ref{IoV}, current IoV drivers can only access traffic or road information from the cockpit display screen, which is quite inconvenient and might bring safety risks.
With VR, these sensed data can be visibly presented on the VR glass or the front car glass.

Some recent work began to take the novel augmented VR technology into the development of IoV systems.
For example, \cite{Qiu2018AVR} broadened the vehicle’s visual horizon by enabling it to share visual information (i.e., 3D features of the environment) with nearby vehicles.
Generally, the design of the augmented VR-empowered IoV system requires novel relative positioning techniques, new perspective transformation methods, approaches to predict the real-time motion of objects, and adaptive streaming strategies \cite{Allmacher2019VR}.
For the monitoring and safety issues, \cite{Alam2017JNCA} presented an augmented VR based IoT prototype system.

Besides the IoV applications, the VR technologies can be applied to establish a drone simulator in a 3D environment.
For example, \cite{Nguyen2019AIVR} developed a prototype DroneVR based on a web application to help operators manipulate the drones, where practical user studies were carried out to improve the designed prototype.
\cite{Choi2000IROS} proposed a distributed virtual environment collaborative simulator to analyze the operating state of underwater unmanned vehicles. In the created virtual environment, vehicles can interact and cooperate with each other, which is helpful in reducing the verification and testing time of underwater unmanned vehicles and solving multi-agent problems.

\subsection{VR for Smart Industry}
The Industry 4.0 is the continuous automation of traditional manufacturing and industrial practices, with IoT and machine-to-machine communication integrated for improved communication and self-monitoring. VR opens up different ways to create smart solutions for the manufacturing industry, which could help to refine the production capability to meet the needs
of customers.
\cite{Frontoni2018CPS} developed a novel Cyber Physical System (CPS) architecture providing real time visualization of complicated industrial process, which integrates realistic virtual models to bridge the gap between design and manufacturing. The results from a real industrial environment indicate that the proposed CPS performs well with improved visibility and utilization, and reduced waste.
\cite{Linn2017VSMM} also proposed a concept to implement the remote inspections of manufacturing machines in order to provide reliable maintenance process for factories, which integrates an anomaly detection module to find problems on the inspected object.

\subsection{VR for Healthcare}
The emerging VR technologies have also attracted the increasing use of healthcare applications. The advantages with VR could improve the quality and efficiency of treatments and accordingly improve the health of the patients. The VR-enabled healthcare industry enables remote monitoring in the healthcare sector, end-to-end connectivity and affordability, personalized health consultation, and empowers physicians to deliver remote medical assistance.
For instance, \cite{Postolache2020JSAC} applies VR serious games and wearable sensor network in physical rehabilitation monitoring to provide personalized exercises for patients and reduce the cost. The proposed system could facilitate physicians with the information about training outcome, and more methodologies can be integrated to improve the performance of the system in the future.
\cite{Hu2017TSMC} developed a cyber physical system for motion recognition and training, which integrated VR to implement interactive virtual scene experience and wireless sensors to monitor the motion training status. The proposed VR system could be potentially used for next-generation rehabilitation due to its automatic motion analysis.

\section{Issues and Future Research Directions}\label{sec_issue}
This section presents issues on constructing VR systems, and introduces future research directions.

\subsection{Issues on System Design}
We first outlook the potential issue on an improved VR system design.

\subsubsection*{Crowdsourced VR Video Generation}
The current VR market lacks of UHD video sources, owing to the harsh demand for UHD 360-degree video acquisition equipment and the long video shooting process. Except for a low-cost design on video shooting devices, the VR content can also be generated from aggression of crowdsoured image or video clips \cite{Wu2016TMC, Wu2017INFOCOM, Wu2017IPSN}. However, such methods often have resource constraints in terms of bandwidth, storage, and processing capability, which limit the number of videos that can be crowdsourced. It is challenging to use limited resources to crowdsource video that best recover the 360-degree target area.

\subsubsection*{Interactive 360-degree Video Telephony System}
POI360, as the first POrtable Interactive 360-degree video telephony system, confronts an intrinsic conflict between the strict latency requirement of interactive video and the highly dynamic cellular network condition \cite{Xie2017CoNEXT_POI360}. The situation is even more challenging when the heavy 360-degree video traffic pushes the cellular link to its limit, as a fluctuating link bandwidth can easily fall below the high traffic load. POI360's rate control must detect this and reduce the traffic rate promptly to prevent video freezes.

\subsubsection*{Lightweight VR Solutions for Wearable IoT Services}
Current wearable-device-based VR solutions are costly and inconvenient to carry. VR HMD has a built-in processor and battery, leading to a heavy HMD to wear. Therefore, current HMDs cannot offer us a lightweight yet high-quality VR experience. For most IoT applications, it is urgently required a lightweight VR solution design.

\subsubsection*{Cross-platform Supportable VR System}
The current application-based VR solutions require installing the specific app in advance and lack cross-platform support (i.e., a VR activity of one IoT application cannot be used in other applications directly).
\cite{Qiao2018WebAR} envisioned that web-based solutions with natural cross-platform advantages is attracting more and more attention and is opening up a new research field of VR system design.

\subsection{Future Research Directions}
There are also some issues concerning the enabling technologies for practical VR applications.

\subsubsection*{User QoE Optimization}
Despite the above mentioned bitrate adaption design, we presents some potential means to improve the user QoE, i.e.,

\begin{itemize}
    \item Reducing the video decoding operation complexity, and thus the processing delay can be reduced.
    \item Minimizing the transmitted content size (\# bits), and thus the transmission delay can be reduced.
    \item Maximizing the transmission bandwidth (bits per second), and thus the transmission delay can be reduced.
    \item Enhancing the viewpoint prediction accuracy, and thus the rebuffering delay can be reduced. Meanwhile, this can also enhance the video quality in the viewport.
\end{itemize}

\subsubsection*{Fine-grained User-centric FoV Prediction}
Among various research efforts towards the VR streaming, head movement prediction is one essential yet daunting task that needs to be addressed urgently. Besides enabling bandwidth-efficient VR streaming, predicting head movement accurately can significantly reduce the motion-to-photon delay since it would be possible to render exactly one viewport from the downloaded video portions in advance before the head moves. More importantly, the content view characteristics are various across different users. It is urgently required to design a personalized FoV prediction model to further improve the VR streaming efficiency.

Here we illustrate some extension ideas for find-grained FoV prediction, including
\begin{itemize}
    \item Adding external information. Besides the 3DoF or 6DoF information, there also exist external information that can be obtained from the sensors on HMD. For example, eye-trackers are installed in some recent HMD, where the eye-tracking information can highly likely to improve the FoV prediction performance.
    \item Exploiting accurate user portrait. Current FoV prediction schemes are based on user portrait from historical view trajectory, which might less accurate.
    \item Brain electrical information might also help improve the FoV prediction accuracy.
\end{itemize}

\subsubsection*{HMD-to-HMD Communication}
Compared to the traditional device-to-device communication, the connection between HMDs is also urgently required but faces much more challenges.
First, the VR contents are usually UHD with high data volume, thus the radio technology should guarantee a high downlink rate.
Moreover, the VR users are always with dynamic movement, thus the signals with new radios (e.g., mmWave or VLC) might be easily blocked and introduce an obvious decline on the received signal strength.

\subsubsection*{Privacy-preserving VR Framework Design}
The VR systems generate, collect, and analyze large volume of data to derive intelligence, privacy becomes a great concern. The privacy issue is caused by several special characteristics of both VR technologies and IoT application requirements. The collection of user dynamic view behaviors and characteristics may pose threats to the privacy of people's preference information.
In VR systems, the multiple-dimensional information leakage might pose users under a more dangerous situation, which will be a matter of continuing concern.

\subsubsection*{Public Dataset Contribution}
Although many communities have contributed some great efforts on public dataset generation as discussed in Section \ref{sec_HMD}, it is still not enough for efficiently mining key characteristics from the limited amounts of user traces.
We need more contribution on public datasets, with which the needs of the research community on VR streaming can be efficiently met.

\section{Conclusion}\label{sec_conclusion}
VR has emerged as one of the most widely adopted technologies for IoT applications. With issues with current server-client architecture (including projection, coding, rendering, streaming, security, and privacy), many have sought for solutions with new system design, dynamic view prediction techniques, efficient processing/rendering techniques and adaptive VR streaming strategies for improved QoE. In particular, we have focused on the realization of VR streaming in detail from the standpoint of existing literature, solutions, and future research and implementation issues. We first overviewed VR streaming architecture along with their characteristics, elaborated various issues and challenges that affect the system performance, then we provided a detailed overview of the HMD, measurement platform and public dataset. Through this survey, we have also presented the state-of-the-art literature regarding the existing VR streaming solutions, architectures and platforms, as well as other aspects of VR systems such as projection schemes, coding/decoding, rendering, FoV prediction, and streaming in VR streaming networks. In the end, we also identified research challenges and issues that may be considered by the research community to design concrete architecture of VR technology for IoT systems. The future of VR streaming in the commercialization is speculative, and we believe that more research is needed to find the common ground for applying VR technology to current IoT systems.

\section*{Acknowledgement}
This work was supported by the National Natural Science Foundation of China under Grants 62072486, U1911201, 61802452, the Natural Science Foundation of Guangdong under Grant 2018A030310079, the project ``PCL Future Greater-Bay Area Network Facilities for Large-scale Experiments and Applications (LZC0019)'', and Guangdong Special Support Program under Grant 2017TX04X148.



\begin{thebibliography}{100}
\providecommand{\url}[1]{#1}
\csname url@samestyle\endcsname
\providecommand{\newblock}{\relax}
\providecommand{\bibinfo}[2]{#2}
\providecommand{\BIBentrySTDinterwordspacing}{\spaceskip=0pt\relax}
\providecommand{\BIBentryALTinterwordstretchfactor}{4}
\providecommand{\BIBentryALTinterwordspacing}{\spaceskip=\fontdimen2\font plus
\BIBentryALTinterwordstretchfactor\fontdimen3\font minus
  \fontdimen4\font\relax}
\providecommand{\BIBforeignlanguage}[2]{{%
\expandafter\ifx\csname l@#1\endcsname\relax
\typeout{** WARNING: IEEEtran.bst: No hyphenation pattern has been}%
\typeout{** loaded for the language `#1'. Using the pattern for}%
\typeout{** the default language instead.}%
\else
\language=\csname l@#1\endcsname
\fi
#2}}
\providecommand{\BIBdecl}{\relax}
\BIBdecl

\bibitem{Statista2019IoT}
Statista, ``{Size of the {Internet of Things} ({IoT}) market worldwide from
  2017 to 2025},''
  \url{https://www.statista.com/statistics/976313/global-iot-market-size/},
  2019.

\bibitem{Afzal2017Characterization}
S.~Afzal, J.~Chen, and K.~K. Ramakrishnan, ``Characterization of 360-degree
  videos,'' in \emph{Proc. of the Workshop on Virtual Reality and Augmented
  Reality Network (VR/AR Network)}, 2017, pp. 1--6.

\bibitem{Xiao2017MM_OpTile}
M.~Xiao, C.~Zhou, Y.~Liu, and S.~Chen, ``{OpTile}: Toward optimal tiling in
  360-degree video streaming,'' in \emph{Proc. of the 25th ACM Multimedia
  Conference (MM)}, 2017, pp. 708--716.

\bibitem{Liu2019TMM}
Y.~{Liu}, J.~{Liu}, A.~{Argyriou}, and S.~{Ci}, ``{MEC}-assisted panoramic {VR}
  video streaming over millimeter wave mobile networks,'' \emph{IEEE
  Transactions on Multimedia}, vol.~21, no.~5, pp. 1302--1316, May 2019.

\bibitem{Xie2017MM_360ProbDASH}
L.~Xie, Z.~Xu, Y.~Ban, X.~Zhang, and Z.~Guo, ``{360ProbDASH}: Improving {QoE}
  of 360 video streaming using tile-based {HTTP} adaptive streaming,'' in
  \emph{Proc. of the ACM on Multimedia Conference (MM)}, 2017, pp. 315--323.

\bibitem{Shi2019MMSys}
S.~Shi, V.~Gupta, M.~Hwang, and R.~Jana, ``Mobile {VR} on edge cloud: A
  latency-driven design,'' in \emph{Proc. of the 10th ACM Multimedia Systems
  Conference (MMSys)}, 2019, pp. 222--231.

\bibitem{Qian2016AllThingsCellular}
F.~Qian, L.~Ji, B.~Han, and V.~Gopalakrishnan, ``Optimizing 360 video delivery
  over cellular networks,'' in \emph{Proc. of the 5th Workshop on All Things
  Cellular: Operations, Applications and Challenges}, 2016, pp. 1--6.

\bibitem{Mangiante2017VRARNetwork}
S.~Mangiante, G.~Klas, A.~Navon, Z.~GuanHua, J.~Ran, and M.~D. Silva, ``{VR} is
  on the edge: How to deliver 360-degree videos in mobile networks,'' in
  \emph{Proc. of the SIGCOMM Workshop on Virtual Reality and Augmented Reality
  Network (VR/AR Network)}, 2017, pp. 30--35.

\bibitem{Sun2018MMSys}
L.~Sun, F.~Duanmu, Y.~Liu, Y.~Wang, Y.~Ye, H.~Shi, and D.~Dai, ``Multi-path
  multi-tier 360-degree video streaming in {5G} networks,'' in \emph{Proc. of
  the 9th ACM Multimedia Systems Conference (MMSys)}, 2018, pp. 162--173.

\bibitem{Abari2017NSDI}
O.~Abari, D.~Bharadia, A.~Duffield, and D.~Katabi, ``Enabling high-quality
  untethered virtual reality,'' in \emph{Proc. of 14th {USENIX} Symposium on
  Networked Systems Design and Implementation ({NSDI})}, Boston, MA, Mar. 2017,
  pp. 531--544.

\bibitem{Zhong2017APSys}
R.~Zhong, M.~Wang, Z.~Chen, L.~Liu, Y.~Liu, J.~Zhang, L.~Zhang, and
  T.~Moscibroda, ``On building a programmable wireless high-quality virtual
  reality system using commodity hardware,'' in \emph{Proc. of the 8th
  Asia-Pacific Workshop on Systems (APSys)}, 2017, pp. 1--7.

\bibitem{Khan2019arXiv}
M.~Khan and J.~Chakareski, ``Visible light communication for next generation
  untethered virtual reality systems,'' \emph{CoRR}, vol. abs/1904.03735, 2019.

\bibitem{HU2017JNCA}
P.~Hu, S.~Dhelim, H.~Ning, and T.~Qiu, ``Survey on fog computing: architecture,
  key technologies, applications and open issues,'' \emph{Journal of Network
  and Computer Applications}, vol.~98, pp. 27 -- 42, 2017.

\bibitem{HSIEH2018JNCA}
H.-C. Hsieh, J.-L. Chen, and A.~Benslimane, ``{5G} virtualized multi-access
  edge computing platform for {IoT} applications,'' \emph{Journal of Network
  and Computer Applications}, vol. 115, pp. 94 -- 102, 2018.

\bibitem{RAY2019JNCA}
P.~P. Ray, D.~Dash, and D.~De, ``Edge computing for internet of things: A
  survey, e-healthcare case study and future direction,'' \emph{Journal of
  Network and Computer Applications}, vol. 140, pp. 1 -- 22, 2019.

\bibitem{ELAZHARY2019JNCA}
H.~Elazhary, ``Internet of things (iot), mobile cloud, cloudlet, mobile iot,
  iot cloud, fog, mobile edge, and edge emerging computing paradigms:
  Disambiguation and research directions,'' \emph{Journal of Network and
  Computer Applications}, vol. 128, pp. 105 -- 140, 2019.

\bibitem{LIU2020JNCA}
L.~Liu, Z.~Lu, L.~Wang, X.~Chen, and X.~Wen, ``Large-volume data dissemination
  for cellular-assisted automated driving with edge intelligence,''
  \emph{Journal of Network and Computer Applications}, vol. 155, p. 102535,
  2020.

\bibitem{Zhang2019NOSSDAV}
L.~Zhang, A.~Sun, R.~Shea, J.~Liu, and M.~Zhang, ``Rendering multi-party mobile
  augmented reality from edge,'' in \emph{Proc. of the 29th ACM Workshop on
  Network and Operating Systems Support for Digital Audio and Video (NOSSDAV)},
  2019, pp. 67--72.

\bibitem{Mahzari2018MM}
A.~Mahzari, A.~Taghavi~Nasrabadi, A.~Samiei, and R.~Prakash, ``{FoV}-aware edge
  caching for adaptive 360 video streaming,'' in \emph{Proc. of ACM Multimedia
  Conference (MM)}, 2018, pp. 173--181.

\bibitem{Hou2018VRARNetwork}
X.~Hou, S.~Dey, J.~Zhang, and M.~Budagavi, ``Predictive view generation to
  enable mobile 360-degree and {VR} experiences,'' in \emph{Proc. of the
  SIGCOMM Workshop on Virtual Reality and Augmented Reality Network}, 2018, pp.
  20--26.

\bibitem{Liu2018MobiSys}
L.~Liu, R.~Zhong, W.~Zhang, Y.~Liu, J.~Zhang, L.~Zhang, and M.~Gruteser,
  ``Cutting the cord: Designing a high-quality untethered {VR} system with low
  latency remote rendering,'' in \emph{Proc. of the 16th Annual International
  Conference on Mobile Systems, Applications, and Services (MobiSys)}, 2018,
  pp. 68--80.

\bibitem{Duanmu2017JSTSP}
Z.~Duanmu, K.~Zeng, K.~Ma, A.~Rehman, and Z.~Wang, ``A quality-of-experience
  index for streaming video,'' \emph{IEEE Journal of Selected Topics in Signal
  Processing}, vol.~11, no.~1, pp. 154--166, Feb. 2017.

\bibitem{DIASDEASSUNCAO2018JNCA}
M.~Assuncao, A.~Veith, and R.~Buyya, ``Distributed data stream processing and
  edge computing: A survey on resource elasticity and future directions,''
  \emph{Journal of Network and Computer Applications}, vol. 103, pp. 1 -- 17,
  2018.

\bibitem{Fernandez2014JNCA}
V.~Fernandez, J.~M. Orduna, and P.~Morillo, ``Server implementations for
  improving the performance of car systems based on mobile phones,''
  \emph{Journal of Network and Computer Applications}, vol.~44, pp. 72 -- 82,
  2014.

\bibitem{Jarschel2014CM}
M.~{Jarschel}, T.~{Zinner}, T.~{Hossfeld}, P.~{Tran-Gia}, and W.~{Kellerer},
  ``Interfaces, attributes, and use cases: A compass for sdn,'' \emph{IEEE
  Communications Magazine}, vol.~52, no.~6, pp. 210--217, 2014.

\bibitem{Lange2019CNSM}
S.~{Lange}, H.~{Kim}, S.~{Jeong}, H.~{Choi}, J.~{Yoo}, and J.~W. {Hong},
  ``Predicting vnf deployment decisions under dynamically changing network
  conditions,'' in \emph{Proc. of 15th International Conference on Network and
  Service Management (CNSM)}, 2019, pp. 1--9.

\bibitem{PlayStationVR}
PlayStation, \url{https://www.playstation.com/en-us/explore/playstation-vr/}.

\bibitem{Lo2017MMSys}
W.-C. Lo, C.-L. Fan, J.~Lee, C.-Y. Huang, K.-T. Chen, and C.-H. Hsu,
  ``360$^\circ$ video viewing dataset in head-mounted virtual reality,'' in
  \emph{Proc. of the ACM on Multimedia Systems Conference (MMSys)}, 2017, pp.
  211--216.

\bibitem{Corbillon2017MMSys}
X.~Corbillon, F.~De~Simone, and G.~Simon, ``360-degree video head movement
  dataset,'' in \emph{Proc. of the 8th ACM on Multimedia Systems Conference
  (MMSys)}, 2017, pp. 199--204.

\bibitem{Wu2017MMSys}
C.~Wu, Z.~Tan, Z.~Wang, and S.~Yang, ``A dataset for exploring user behaviors
  in {VR} spherical video streaming,'' in \emph{Proc. of the 8th ACM on
  Multimedia Systems Conference (MMSys)}, 2017, pp. 193--198.

\bibitem{David2018MMSys}
E.~J. David, J.~Guti{\'e}rrez, A.~Coutrot, M.~P. Da~Silva, and P.~L. Callet,
  ``A dataset of head and eye movements for 360$^\circ$ videos,'' in
  \emph{Proc. of the 9th ACM Multimedia Systems Conference (MMSys)}, 2018, pp.
  432--437.

\bibitem{Ozcinar2018QoMEX}
O.~Cagri and S.~Aljosa, ``Visual attention in omnidirectional video for virtual
  reality applications,'' in \emph{Proc. of Tenth International Conference on
  Quality of Multimedia Experience (QoMEX)}, 2018, pp. 1--6.

\bibitem{Duanmu2018ICME}
F.~{Duanmu}, Y.~{Mao}, S.~{Liu}, S.~{Srinivasan}, and Y.~{Wang}, ``A subjective
  study of viewer navigation behaviors when watching 360-degree videos on
  computers,'' in \emph{Proc. of IEEE International Conference on Multimedia
  and Expo (ICME)}, July 2018, pp. 1--6.

\bibitem{Fremerey2018MMSys_AVTrack}
S.~Fremerey, A.~Singla, K.~Meseberg, and A.~Raake, ``{AVTrack360}: An open
  dataset and software recording people's head rotations watching 360$^\circ$
  videos on an {HMD},'' in \emph{Proc. of the 9th ACM Multimedia Systems
  Conference (MMSys)}, 2018, pp. 403--408.

\bibitem{Nasrabadi2019MMSys}
A.~T. Nasrabadi, A.~Samiei, A.~Mahzari, R.~P. McMahan, R.~Prakash, M.~C.~Q.
  Farias, and M.~M. Carvalho, ``A taxonomy and dataset for 360$^\circ$
  videos,'' in \emph{Proc. of the 10th ACM Multimedia Systems Conference
  (MMSys)}, 2019, pp. 273--278.

\bibitem{Rai2017MMSys}
Y.~Rai, J.~Guti{\'e}rrez, and P.~Le~Callet, ``A dataset of head and eye
  movements for 360 degree images,'' in \emph{Proc. of the 8th ACM on
  Multimedia Systems Conference (MMSys)}, 2017, pp. 205--210.

\bibitem{Xu2018CVPR}
Y.~{Xu}, Y.~{Dong}, J.~{Wu}, Z.~{Sun}, Z.~{Shi}, J.~{Yu}, and S.~{Gao}, ``Gaze
  prediction in dynamic 360° immersive videos,'' in \emph{Proc. of IEEE
  Conference on Computer Vision and Pattern Recognition (CVPR)}, 2018, pp.
  5333--5342.

\bibitem{Fan2017NOSSDAV}
C.-L. Fan, J.~Lee, W.-C. Lo, C.-Y. Huang, K.-T. Chen, and C.-H. Hsu, ``Fixation
  prediction for 360$^\circ$ video streaming in head-mounted virtual reality,''
  in \emph{Proc. of the 27th Workshop on Network and Operating Systems Support
  for Digital Audio and Video (NOSSDAV)}, 2017, pp. 67--72.

\bibitem{He2018MobiSys_Rubiks}
J.~He, M.~A. Qureshi, L.~Qiu, J.~Li, F.~Li, and L.~Han, ``Rubiks: Practical
  360-degree streaming for smartphones,'' in \emph{Proc. of the 16th Annual
  International Conference on Mobile Systems, Applications, and Services
  (MobiSys)}, 2018, pp. 482--494.

\bibitem{Qian2018Flare}
F.~Qian, B.~Han, Q.~Xiao, and V.~Gopalakrishnan, ``Flare: Practical
  viewport-adaptive 360-degree video streaming for mobile devices,'' in
  \emph{Proc. of the Annual International Conference on Mobile Computing and
  Networking (MobiCom)}, 2018, pp. 99--114.

\bibitem{Zhang2019INFOCOM_DRL360}
Y.~Zhang, P.~Zhao, K.~Bian, Y.~Liu, L.~Song, and X.~Li, ``{DRL360}: 360-degree
  video streaming with deep reinforcement learning,'' in \emph{Proc. of IEEE
  Conference on Computer Communications (INFOCOM)}, 2019, pp. 1252--1260.

\bibitem{Nguyen2018MM}
A.~Nguyen, Z.~Yan, and K.~Nahrstedt, ``Your attention is unique: Detecting
  360-degree video saliency in head-mounted display for head movement
  prediction,'' in \emph{Proc. of the 26th ACM International Conference on
  Multimedia (MM)}, 2018, pp. 1190--1198.

\bibitem{Lee2019NOSSDAV}
S.~Lee, D.~Jang, J.~Jeong, and E.-S. Ryu, ``Motion-constrained tile set based
  360-degree video streaming using saliency map prediction,'' in \emph{Proc. of
  the 29th ACM Workshop on Network and Operating Systems Support for Digital
  Audio and Video}, 2019, pp. 20--24.

\bibitem{Xie2018MM_CLS}
L.~Xie, X.~Zhang, and Z.~Guo, ``{CLS}: A cross-user learning based system for
  improving {QoE} in 360-degree video adaptive streaming,'' in \emph{Proc. of
  the 26th ACM International Conference on Multimedia (MM)}, 2018, pp.
  564--572.

\bibitem{LR_DeepLearning}
I.~Goodfellow, Y.~Bengio, and A.~Courville, \emph{Deep learning}.\hskip 1em
  plus 0.5em minus 0.4em\relax MIT press, 2016.

\bibitem{DBSCAN}
M.~Ester, H.-P. Kriegel, J.~Sander, and X.~Xu, ``A density-based algorithm for
  discovering clusters a density-based algorithm for discovering clusters in
  large spatial databases with noise,'' in \emph{Proc. of the Second
  International Conference on Knowledge Discovery and Data Mining (KDD)}, 1996,
  pp. 226--231.

\bibitem{Lee2020MobiCom}
K.~Lee, J.~Yi, Y.~Lee, S.~Choi, and Y.~M. Kim, ``{GROOT}: A real-time streaming
  system of high-fidelity volumetric videos,'' in \emph{Proceedings of the 26th
  Annual International Conference on Mobile Computing and Networking
  (MobiCom)}, 2020.

\bibitem{Clemm2020CM}
A.~{Clemm}, M.~T. {Vega}, H.~K. {Ravuri}, T.~{Wauters}, and F.~D. {Turck},
  ``Toward truly immersive holographic-type communication: Challenges and
  solutions,'' \emph{IEEE Communications Magazine}, vol.~58, no.~1, pp. 93--99,
  2020.

\bibitem{Alsharif2020Symmetry}
M.~H. Alsharif, A.~H. Kelechi, M.~A. Albreem, S.~A. Chaudhry, M.~S. Zia, and
  S.~Kim, ``Sixth generation ({6G}) wireless networks: Vision, research
  activities, challenges and potential solutions,'' \emph{Symmetry}, vol.~12,
  no.~4, p. 676, 2020.

\bibitem{Wakunami2016NatureComm}
K.~Wakunami, P.-Y. Hsieh, R.~Oi, T.~Senoh, H.~Sasaki, Y.~Ichihashi, M.~Okui,
  Y.-P. Huang, and K.~Yamamoto, ``Projection-type see-through holographic
  three-dimensional display,'' \emph{Nature communications}, vol.~7, no.~1, pp.
  1--7, 2016.

\bibitem{Calvanese2019VTM}
E.~{Calvanese Strinati}, S.~{Barbarossa}, J.~L. {Gonzalez-Jimenez},
  D.~{Ktenas}, N.~{Cassiau}, L.~{Maret}, and C.~{Dehos}, ``6g: The next
  frontier: From holographic messaging to artificial intelligence using
  subterahertz and visible light communication,'' \emph{IEEE Vehicular
  Technology Magazine}, vol.~14, no.~3, pp. 42--50, 2019.

\bibitem{Zhou2017MMSys}
C.~Zhou, Z.~Li, and Y.~Liu, ``A measurement study of oculus 360 degree video
  streaming,'' in \emph{Proc. of the 8th ACM Multimedia Systems Conference
  (MMSys)}, 2017, pp. 27--37.

\bibitem{FFmpeg}
FFmpeg, ``Ffmpeg,'' \url{http://www.ffmpeg.org/}.

\bibitem{x264}
x264, ``x264,'' \url{https://x264.org/}.

\bibitem{Feuvre2016MMSys}
J.~Le~Feuvre and C.~Concolato, ``Tiled-based adaptive streaming using
  {MPEG-DASH},'' in \emph{Proc. of the 7th ACM Multimedia Systems Conference
  (MMSys)}, 2016, p.~41.

\bibitem{Hosseini2017VR}
M.~Hosseini, ``View-aware tile-based adaptations in 360 virtual reality video
  streaming,'' in \emph{Proc. of IEEE Virtual Reality (VR)}, 2017, pp.
  423--424.

\bibitem{Niamut2016MMSys}
O.~A. Niamut, E.~Thomas, L.~D'Acunto, C.~Concolato, F.~Denoual, and S.~Y. Lim,
  ``{MPEG DASH SRD}: spatial relationship description,'' in \emph{Proc. of the
  7th ACM Multimedia Systems Conference (MMSys)}, 2016, p.~5.

\bibitem{Zhou2018INFOCOM_ClusTile}
C.~Zhou, M.~Xiao, and Y.~Liu, ``{ClusTile}: Toward minimizing bandwidth in
  360-degree video streaming,'' in \emph{Proc. of IEEE Conference on Computer
  Communications (INFOCOM)}, 2018, pp. 962--970.

\bibitem{Xiao2018MM_miniView}
M.~Xiao, S.~Wang, C.~Zhou, L.~Liu, Z.~Li, Y.~Liu, and S.~Chen, ``{MiniView}
  layout for bandwidth-efficient 360-degree video,'' in \emph{Proc. of ACM
  Multimedia Conference (MM)}, 2018, pp. 914--922.

\bibitem{Guan2019Pano}
Y.~Guan, C.~Zheng, X.~Zhang, Z.~Guo, and J.~Jiang, ``Pano: Optimizing
  360-degree video streaming with a better understanding of quality
  perception,'' in \emph{Proc. of the ACM Special Interest Group on Data
  Communication (SIGCOMM)}, 2019, pp. 394--407.

\bibitem{Zare2016MM}
A.~Zare, A.~Aminlou, M.~M. Hannuksela, and M.~Gabbouj, ``{HEVC}-compliant
  tile-based streaming of panoramic video for virtual reality applications,''
  in \emph{Proc. of the 24th ACM International Conference on Multimedia (MM)},
  2016, p. 601–605.

\bibitem{Liu2019MobiCom}
L.~Liu, H.~Li, and M.~Gruteser, ``Edge assisted real-time object detection for
  mobile augmented reality,'' in \emph{Proc. of the 25th Annual International
  Conference on Mobile Computing and Networking (MobiCom)}, 2019, pp. 1--16.

\bibitem{H.264}
H.264, ``H.264,'' \url{https://www.itu.int/rec/T-REC-H.264}, 2019.

\bibitem{H.265}
H.265, ``H.265,'' \url{https://www.itu.int/rec/T-REC-H.265}, 2019.

\bibitem{Schwarz2007SVC}
H.~Schwarz, D.~Marpe, and T.~Wiegand, ``Overview of the scalable video coding
  extension of the {H.264/AVC} standard,'' \emph{IEEE Transactions on Circuits
  and Systems for Video Technology}, pp. 1--12, 2007.

\bibitem{graziosi_nakagami_kuma_zaghetto_suzuki_tabatabai_2020}
D.~Graziosi, O.~Nakagami, S.~Kuma, A.~Zaghetto, T.~Suzuki, and A.~Tabatabai,
  ``An overview of ongoing point cloud compression standardization activities:
  video-based (v-pcc) and geometry-based (g-pcc),'' \emph{APSIPA Transactions
  on Signal and Information Processing}, vol.~9, pp. 1--13, 2020.

\bibitem{Corbillon2017ICC}
X.~{Corbillon}, G.~{Simon}, A.~{Devlic}, and J.~{Chakareski},
  ``Viewport-adaptive navigable 360-degree video delivery,'' in \emph{Proc. of
  IEEE International Conference on Communications (ICC)}, May 2017, pp. 1--7.

\bibitem{ImageMagick}
ImageMagick, \url{https://imagemagick.org/index.php}, 1999.

\bibitem{GPAC}
GPAC, ``{GPAC},'' \url{https://gpac.wp.imt.fr/}.

\bibitem{EAC}
Google, ``Bringing pixels front and center in {VR} video,''
  \url{https://blog.google/products/google-vr/bringing-pixels-front-and-center-vr-video/},
  2018.

\bibitem{Fu2009TMM}
C.~{Fu}, L.~{Wan}, T.~{Wong}, and C.~{Leung}, ``The rhombic dodecahedron map:
  An efficient scheme for encoding panoramic video,'' \emph{IEEE Transactions
  on Multimedia}, vol.~11, no.~4, pp. 634--644, June 2009.

\bibitem{Graf2017MMSys}
M.~Graf, C.~Timmerer, and C.~Mueller, ``Towards bandwidth efficient adaptive
  streaming of omnidirectional video over {HTTP}: Design, implementation, and
  evaluation,'' in \emph{Proc. of the 8th ACM on Multimedia Systems Conference
  (MMSys)}, 2017, pp. 261--271.

\bibitem{Petrangeli2017MM}
S.~Petrangeli, V.~Swaminathan, M.~Hosseini, and F.~De~Turck, ``An
  {HTTP/2-based} adaptive streaming framework for 360 virtual reality videos,''
  in \emph{Proc. of the 25th ACM Multimedia Conference (MM)}, 2017, pp.
  306--314.

\bibitem{CostaFilho2018MMSys}
R.~I.~T. da~Costa~Filho, M.~C. Luizelli, M.~T. Vega, J.~van~der Hooft,
  S.~Petrangeli, T.~Wauters, F.~De~Turck, and L.~P. Gaspary, ``Predicting the
  performance of virtual reality video streaming in mobile networks,'' in
  \emph{Proc. of ACM Multimedia Systems Conference (MMSys)}, 2018, pp.
  270--283.

\bibitem{Jiang2018LCN}
X.~Jiang, Y.-H. Chiang, Y.~Zhao, and Y.~Ji, ``Plato: Learning-based adaptive
  streaming of 360-degree videos,'' in \emph{Proc. of IEEE 43rd Conference on
  Local Computer Networks (LCN)}, Oct. 2018, pp. 393--400.

\bibitem{Xiao2019ACM_TURC}
G.~Xiao, X.~Chen, M.~Wu, and Z.~Zhou, ``Deep reinforcement learning-driven
  intelligent panoramic video bitrate adaptation,'' in \emph{Proc. of the ACM
  Turing Celebration Conference-China (TURC)}, 2019, p.~41.

\bibitem{Kan2019ICASSP}
N.~Kan, J.~Zou, K.~Tang, C.~Li, N.~Liu, and H.~Xiong, ``Deep reinforcement
  learning-based rate adaptation for adaptive 360-degree video streaming,'' in
  \emph{Proc. of IEEE International Conference on Acoustics, Speech and Signal
  Processing (ICASSP)}, 2019, pp. 4030--4034.

\bibitem{Chen2020NOSSDAV_SR360}
J.~Chen, M.~Hu, Z.~Luo, Z.~Wang, and D.~Wu, ``Sr360: Boosting 360-degree video
  streaming with super-resolution,'' in \emph{Proc. of the 30th ACM Workshop on
  Network and Operating Systems Support for Digital Audio and Video}, 2020.

\bibitem{Kua2017CST}
J.~{Kua}, G.~{Armitage}, and P.~{Branch}, ``A survey of rate adaptation
  techniques for dynamic adaptive streaming over {HTTP},'' \emph{IEEE
  Communications Surveys Tutorials}, vol.~19, no.~3, pp. 1842--1866,
  thirdquarter 2017.

\bibitem{Baldi2000ToN}
M.~{Baldi} and Y.~{Ofek}, ``End-to-end delay analysis of video conferencing
  over packet-switched networks,'' \emph{IEEE/ACM Transactions on Networking},
  vol.~8, no.~4, pp. 479--492, 2000.

\bibitem{Bovy2002PAM}
C.~Bovy, H.~Mertodimedjo, G.~Hooghiemstra, H.~Uijterwaal, and P.~Van~Mieghem,
  ``Analysis of end-to-end delay measurements in internet,'' in \emph{Proc. of
  the Passive and Active Measurement Workshop (PAM)}, vol. 2002, 2002.

\bibitem{Erbad2010MMSys}
A.~Erbad, M.~Tayarani~Najaran, and C.~Krasic, ``Paceline: Latency management
  through adaptive output,'' in \emph{Proceedings of the First Annual ACM SIGMM
  Conference on Multimedia Systems (MMSys)}.\hskip 1em plus 0.5em minus
  0.4em\relax New York, NY, USA: Association for Computing Machinery, 2010, pp.
  181--192.

\bibitem{Kumar2017RTSS}
R.~{Kumar}, M.~{Hasan}, S.~{Padhy}, K.~{Evchenko}, L.~{Piramanayagam},
  S.~{Mohan}, and R.~B. {Bobba}, ``End-to-end network delay guarantees for
  real-time systems using {SDN},'' in \emph{Proc. of IEEE Real-Time Systems
  Symposium (RTSS)}, 2017, pp. 231--242.

\bibitem{Clarity2014WhitePaper}
V.~Clarity, ``A white paper: Understanding {MOS}, {JND} and {PSNR},'' 2014.

\bibitem{ANWAR2020SCIENCECHINA}
M.~S. ANWAR, J.~WANG, A.~ULLAH, W.~KHAN, S.~AHMAD, and Z.~FEI, ``Measuring
  quality of experience for 360-degree videos in virtual reality,''
  \emph{SCIENCE CHINA Information Sciences}, vol.~63, no.~10, 2020.

\bibitem{Schatz2017QoEMX}
R.~{Schatz}, A.~{Sackl}, C.~{Timmerer}, and B.~{Gardlo}, ``Towards subjective
  quality of experience assessment for omnidirectional video streaming,'' in
  \emph{2017 Ninth International Conference on Quality of Multimedia Experience
  (QoMEX)}, 2017, pp. 1--6.

\bibitem{Tran2017MMSP}
H.~T.~T. {Tran}, N.~P. {Ngoc}, C.~T. {Pham}, Y.~J. {Jung}, and T.~C. {Thang},
  ``A subjective study on qoe of 360 video for vr communication,'' in
  \emph{Proc. of IEEE 19th International Workshop on Multimedia Signal
  Processing (MMSP)}, 2017, pp. 1--6.

\bibitem{Anwar2020Access}
M.~S. {Anwar}, J.~{Wang}, W.~{Khan}, A.~{Ullah}, S.~{Ahmad}, and Z.~{Fei},
  ``Subjective qoe of 360-degree virtual reality videos and machine learning
  predictions,'' \emph{IEEE Access}, vol.~8, pp. 148\,084--148\,099, 2020.

\bibitem{Corbillon2017MM}
X.~Corbillon, A.~Devlic, G.~Simon, and J.~Chakareski, ``Optimal set of
  360-degree videos for viewport-adaptive streaming,'' in \emph{Proc. of the
  25th ACM International Conference on Multimedia (MM)}, 2017, pp. 943--951.

\bibitem{Xiao2018INFOCOM}
M.~{Xiao}, C.~{Zhou}, V.~{Swaminathan}, Y.~{Liu}, and S.~{Chen}, ``{BAS-360}:
  Exploring spatial and temporal adaptability in 360-degree videos over
  {HTTP/2},'' in \emph{Proc. of IEEE Conference on Computer Communications
  (INFOCOM)}, Apr. 2018, pp. 953--961.

\bibitem{Xu2019TPAMI}
M.~{Xu}, Y.~{Song}, J.~{Wang}, M.~{Qiao}, L.~{Huo}, and Z.~{Wang}, ``Predicting
  head movement in panoramic video: A deep reinforcement learning approach,''
  \emph{IEEE Transactions on Pattern Analysis and Machine Intelligence}, pp.
  1--14, 2019.

\bibitem{Yi2019NOSSDAV}
J.~Yi, S.~Luo, and Z.~Yan, ``A measurement study of youtube 360-degree live
  video streaming,'' in \emph{Proc. of the 29th ACM Workshop on Network and
  Operating Systems Support for Digital Audio and Video (NOSSDAV)}, 2019, pp.
  49--54.

\bibitem{VandenBroeck2017MM}
M.~V.~d. Broeck, F.~Kawsar, and J.~Sch\"{o}ning, ``It's all around you:
  Exploring 360$^\circ$ video viewing experiences on mobile devices,'' in
  \emph{Proc. of the 25th ACM International Conference on Multimedia (MM)},
  2017, pp. 762--768.

\bibitem{Li2018MM}
C.~Li, M.~Xu, X.~Du, and Z.~Wang, ``Bridge the gap between {VQA} and human
  behavior on omnidirectional video: A large-scale dataset and a deep learning
  model,'' in \emph{Proc. of the 26th ACM International Conference on
  Multimedia (MM)}, 2018, pp. 932--940.

\bibitem{Xue2014ICMEW}
J.~Xue, D.-Q. Zhang, H.~Yu, and C.~W. Chen, ``Assessing quality of experience
  for adaptive {HTTP} video streaming,'' in \emph{Proc. of IEEE International
  Conference on Multimedia and Expo Workshops (ICMEW)}, July 2014, pp. 1--6.

\bibitem{Yu2015ISMAR}
M.~{Yu}, H.~{Lakshman}, and B.~{Girod}, ``A framework to evaluate
  omnidirectional video coding schemes,'' in \emph{Proc. of IEEE International
  Symposium on Mixed and Augmented Reality (ISMAR)}, Sept. 2015, pp. 31--36.

\bibitem{Wang2003SSIM}
Z.~{Wang}, E.~P. {Simoncelli}, and A.~C. {Bovik}, ``Multiscale structural
  similarity for image quality assessment,'' in \emph{Proc. of the
  Thrity-Seventh Asilomar Conference on Signals, Systems Computers}, vol.~2,
  no.~2, Nov. 2003, pp. 1398--1402.

\bibitem{Chou1995PSPNR}
{Chun-Hsien Chou} and {Yun-Chin Li}, ``A perceptually tuned subband image coder
  based on the measure of just-noticeable-distortion profile,'' \emph{IEEE
  Transactions on Circuits and Systems for Video Technology}, vol.~5, no.~6,
  pp. 467--476, Dec. 1995.

\bibitem{Zhao2011TCSVT}
Y.~{Zhao}, L.~{Yu}, Z.~{Chen}, and C.~{Zhu}, ``Video quality assessment based
  on measuring perceptual noise from spatial and temporal perspectives,''
  \emph{IEEE Transactions on Circuits and Systems for Video Technology},
  vol.~21, no.~12, pp. 1890--1902, Dec. 2011.

\bibitem{sector2012series}
S.~SECTOR and O.~ITU, ``Series y: Global information infrastructure, internet
  protocol aspects and next-generation networks next generation
  networks--frameworks and functional architecture models,''
  \emph{International Telecommunication Union, Geneva, Switzerland,
  Recommendation ITU-T Y}, vol. 2060, 2012.

\bibitem{Song2020TMM}
J.~{Song}, F.~{Yang}, W.~{Zhang}, W.~{Zou}, Y.~{Fan}, and P.~{Di}, ``A fast
  {FoV}-switching {DASH} system based on tiling mechanism for practical
  omnidirectional video services,'' \emph{IEEE Transactions on Multimedia}, pp.
  1--12, 2020.

\bibitem{Xie2017CoNEXT_POI360}
X.~Xie and X.~Zhang, ``{POI360}: Panoramic mobile video telephony over {LTE}
  cellular networks,'' in \emph{Proc. of the 13th International Conference on
  Emerging Networking EXperiments and Technologies (CoNEXT)}, 2017, pp.
  336--349.

\bibitem{Slavova2018VR}
Y.~{Slavova} and M.~{Mu}, ``A comparative study of the learning outcomes and
  experience of {VR} in education,'' in \emph{Proc. of IEEE Conference on
  Virtual Reality and 3D User Interfaces (VR)}, Mar. 2018, pp. 685--686.

\bibitem{Kim2019VR}
H.~{Kim}, S.~{Nah}, J.~{Oh}, and H.~{Ryu}, ``{VR-MOOCs}: A learning management
  system for {VR} education,'' in \emph{Proc. of IEEE Conference on Virtual
  Reality and 3D User Interfaces (VR)}, Mar. 2019, pp. 1325--1326.

\bibitem{Wei2019VR}
H.~{Wei}, Y.~{Liu}, and Y.~{Wang}, ``Building {AR}-based optical experiment
  applications in a {VR} course,'' in \emph{Proc. of IEEE Conference on Virtual
  Reality and 3D User Interfaces (VR)}, Mar. 2019, pp. 1225--1226.

\bibitem{Fiederer2019VR}
L.~D.~J. {Fiederer}, H.~{Alwanni}, M.~{Völker}, O.~{Schnell}, J.~{Beck}, and
  T.~{Ball}, ``A research framework for virtual-reality neurosurgery based on
  open-source tools,'' in \emph{Proc. of IEEE Conference on Virtual Reality and
  3D User Interfaces (VR)}, Mar. 2019, pp. 922--924.

\bibitem{Todsen2018VR}
T.~{Todsen}, J.~{Melchiors}, and K.~{Wennerwaldt}, ``Use of virtual reality to
  teach teamwork and patient safety in surgical education,'' in \emph{Proc. of
  IEEE Conference on Virtual Reality and 3D User Interfaces (VR)}, Mar. 2018,
  pp. 1--1.

\bibitem{Mertz2019Pulse}
L.~{Mertz}, ``Virtual reality pioneer tom furness on the past, present, and
  future of {VR} in health care,'' \emph{IEEE Pulse}, vol.~10, no.~3, pp.
  9--11, May 2019.

\bibitem{Banks2004JNCA}
J.~Banks, G.~Ericksson, K.~Burrage, P.~Yellowlees, S.~Ivermee, and J.~Tichon,
  ``Constructing the hallucinations of psychosis in virtual reality,''
  \emph{Journal of Network and Computer Applications}, vol.~27, no.~1, pp. 1 --
  11, 2004.

\bibitem{IoVMarket}
AppliedMarketResearch,
  \url{https://www.alliedmarketresearch.com/internet-of-vehicles-market}, 2018.

\bibitem{Qiu2018AVR}
H.~Qiu, F.~Ahmad, F.~Bai, M.~Gruteser, and R.~Govindan, ``{AVR}: Augmented
  vehicular reality,'' in \emph{Proc. of the 16th Annual International
  Conference on Mobile Systems, Applications, and Services (MobiSys)}, 2018,
  pp. 81--95.

\bibitem{Allmacher2019VR}
C.~{Allmacher}, M.~{Dudczig}, S.~{Knopp}, and P.~{Klimant}, ``Virtual reality
  for virtual commissioning of automated guided vehicles,'' in \emph{Proc. of
  IEEE Conference on Virtual Reality and 3D User Interfaces (VR)}, Mar. 2019,
  pp. 838--839.

\bibitem{Alam2017JNCA}
F.~Alam, S.~Katsikas, O.~Beltramello, and S.~Hadjiefthymiades, ``Augmented and
  virtual reality based monitoring and safety system: A prototype iot
  platform,'' \emph{Journal of Network and Computer Applications}, vol.~89, pp.
  109 -- 119, 2017.

\bibitem{Nguyen2019AIVR}
V.~T. {Nguyen}, K.~{Jung}, and T.~{Dang}, ``Dronevr: A web virtual reality
  simulator for drone operator,'' in \emph{2019 IEEE International Conference
  on Artificial Intelligence and Virtual Reality (AIVR)}, 2019, pp. 257--2575.

\bibitem{Choi2000IROS}
S.~K. {Choi}, S.~A. {Menor}, and J.~{Yuh}, ``Distributed virtual environment
  collaborative simulator for underwater robots,'' in \emph{Proceedings. 2000
  IEEE/RSJ International Conference on Intelligent Robots and Systems (IROS)},
  vol.~2, 2000, pp. 861--866.

\bibitem{Frontoni2018CPS}
E.~Frontoni, J.~Loncarski, R.~Pierdicca, M.~Bernardini, and M.~Sasso, ``Cyber
  physical systems for industry 4.0: Towards real time virtual reality in smart
  manufacturing,'' in \emph{Augmented Reality, Virtual Reality, and Computer
  Graphics}, L.~T. De~Paolis and P.~Bourdot, Eds.\hskip 1em plus 0.5em minus
  0.4em\relax Cham: Springer International Publishing, 2018, pp. 422--434.

\bibitem{Linn2017VSMM}
C.~{Linn}, S.~{Bender}, J.~{Prosser}, K.~{Schmitt}, and D.~{Werth}, ``Virtual
  remote inspection — a new concept for virtual reality enhanced real-time
  maintenance,'' in \emph{Proc. of 23rd International Conference on Virtual
  System Multimedia (VSMM)}, 2017, pp. 1--6.

\bibitem{Postolache2020JSAC}
O.~{Postolache}, R.~{Alexandre}, O.~{Geman}, D.~{Jude Hemanth}, D.~{Gupta}, and
  A.~{Khanna}, ``Remote monitoring of physical rehabilitation of stroke
  patients using iot and virtual reality,'' \emph{IEEE Journal on Selected
  Areas in Communications}, pp. 1--12, 2020.

\bibitem{Hu2017TSMC}
F.~{Hu}, Q.~{Hao}, Q.~{Sun}, X.~{Cao}, R.~{Ma}, T.~{Zhang}, Y.~{Patil}, and
  J.~{Lu}, ``Cyberphysical system with virtual reality for intelligent motion
  recognition and training,'' \emph{IEEE Transactions on Systems, Man, and
  Cybernetics: Systems}, vol.~47, no.~2, pp. 347--363, 2017.

\bibitem{Wu2016TMC}
Y.~{Wu}, Y.~{Wang}, W.~{Hu}, and G.~{Cao}, ``{SmartPhoto}: A resource-aware
  crowdsourcing approach for image sensing with smartphones,'' \emph{IEEE
  Transactions on Mobile Computing}, vol.~15, no.~5, pp. 1249--1263, May 2016.

\bibitem{Wu2017INFOCOM}
Y.~{Wu}, Y.~{Wang}, and G.~{Cao}, ``Photo crowdsourcing for area coverage in
  resource constrained environments,'' in \emph{Proc. of IEEE Conference on
  Computer Communications (INFOCOM)}, May 2017, pp. 1--9.

\bibitem{Wu2017IPSN}
Y.~{Wu} and G.~{Cao}, ``{VideoMec}: A metadata-enhanced crowdsourcing system
  for mobile videos,'' in \emph{Proc. of 16th ACM/IEEE International Conference
  on Information Processing in Sensor Networks (IPSN)}, Apr. 2017, pp.
  143--154.

\bibitem{Qiao2018WebAR}
X.~{Qiao}, P.~{Ren}, S.~{Dustdar}, and J.~{Chen}, ``A new era for web {AR} with
  mobile edge computing,'' \emph{IEEE Internet Computing}, vol.~22, no.~4, pp.
  46--55, July 2018.

\end{thebibliography}
\end{document}